\documentclass[fleqn,10pt]{wlscirep}
\usepackage[utf8]{inputenc}
\usepackage[T1]{fontenc}
\pdfoutput=1

\usepackage{makecell}

\usepackage{float}
\usepackage{fixltx2e}
\usepackage{lineno}
\usepackage{graphicx}
\usepackage{gensymb}
\usepackage{nameref}
\usepackage{amssymb}
\usepackage{caption}
\usepackage{subcaption}
\usepackage{comment}
\usepackage{array}

\title{Reconstruction of Excitation Waves from Mechanical Deformation using Physics-Informed Neural Networks}

\author[1,2,*]{Nathan Dermul}
\author[1,2]{Hans Dierckx}
\affil[1]{KU Leuven, Department of Mathematics, Kortrijk, 8500, Belgium}
\affil[2]{KU Leuven, iSi Health, Institute of Physics-based Modeling for In Silico Health: 3000 Leuven, Belgium}

\affil[*]{nathan.dermul@kuleuven.be}

\begin{abstract}
Non-invasive assessment of the electrical activation pattern can significantly contribute to the diagnosis and treatment of cardiac arrhythmias, due to faster and safer diagnosis, improved surgical planning and easier follow-up. One promising path is to measure the mechanical contraction via echocardiography and utilize this as an indirect way of measuring the original activation pattern. To solve this demanding inversion task, we make use of physics-informed neural networks, an upcoming methodology to solve forward and inverse physical problems governed by partial differential equations. In this study, synthetic data sets were created, consisting of 2D excitation waves coupled to an isotropic and linearly deforming elastic medium. We show that for both focal and spiral patterns, the underlying excitation waves can be reconstructed accurately. We test the robustness of the method against Gaussian noise, reduced spatial resolution and projected tri-planar data. In situations where the data quality is heavily reduced, we show how to improve the reconstruction by additional regularization on the wave speed. Results on the optimization of hyperparameters are also discussed. Our findings suggest that physics-informed neural networks hold the potential to solve sparse and noisy bio-mechanical inversion problems and may offer a pathway to non-invasive assessment of certain cardiac arrhythmias. 
\end{abstract}

\begin{document}

\flushbottom
\maketitle

\thispagestyle{empty}

\section*{Introduction}


Heart rhythm disorders affect millions of people. The most prevalent arrhythmia, atrial fibrillation, is a worldwide epidemic striking 33 million people\cite{Rahman2014}. Arrhythmias which affect the lower cardiac chambers, or ventricles, significantly reduce the pumping efficiency of the heart, and account for $80\,\%$ of sudden cardiac deaths, i.e. $12\,\%$ of all deaths worldwide\cite{Mehra2007}. A crucial part of diagnosis and treatment is to image the electrical wave patterns causing the arrhythmia. At present, the most reliable clinical mapping of activation patterns is done during an ablation procedure by inserting a catheter inside the cardiac chambers and making direct contact between the electrode tip and the myocardium. Even though this technique is routinely used in the clinic, important drawbacks are the significant time inside a costly specialized lab environment and the potential risks for the patient due to longer operational procedures. In addition, the thickness of the cardiac walls cannot be ignored in some cases, which reduces the measurements to a surface projection of the real three-dimensional wave. Different solutions have been proposed to measure the electrical pattern non-invasively. Such approaches have the potential to improve pre-surgical diagnosis and planning, while also opening the door towards fully non-invasive diagnosis and treatment, e.g. via external ablation\cite{Graeff2018}.\\

One option for non-invasive mapping of cardiac activity is electrocardiographic imaging (ECGi), which samples the body surface potential in many points and aims to infer the electrical pattern through physical models\cite{Ramanathan2004,Pereira2020}. However, the inverse problem is extremely ill-posed, requiring additional knowledge of the torso structure as well as a significant amount of assumptions about the signal. As such, ECGi is not yet able to precisely and consistently localize focal or re-entrant sources \cite{Duchateau2019}. An alternative non-invasive approach lies in advanced ultrasound imaging of the heart. Herein, one traces the effect of excitation, i.e. the displacement and deformation of the muscle, to infer the spatiotemporal activation sequence \cite{Grondin2019,Grubb2020,Christoph2018}. This approach opens up new possibilities for safe and fast imaging. Furthermore, high frame-rate 3D ultrasound technology has seen important improvements in recent years \cite{Orlowska2020,Tanter2014,Lang2018}. The advantage is that deformation can be directly measured in the heart. The drawback is that arrhythmias are caused by electrical signals, and the mechanical contraction is a consequence of it. Therefore, the precise localization of sources and reconstruction of intricate patterns requires a non-trivial step of inverting the electrical pattern from the mechanical deformation.\\ 

For relatively simple activation patterns, mechanical local activation times (LAT) from strain and strain-rate curves can be measured and interpreted as a proxy for the electrical LAT\cite{Grondin2019}. Over the last ten years, this method of Electromechanical Wave Imaging (EWI) has been tested in silico and in vitro and has been used in clinical trials\cite{Grubb2020}. Note that more complex patterns and non-homogeneous electro-mechanical coupling may cause a difference in the activation sequences recorded via either electrical or mechanical means. It was thereafter shown ex vivo that during ventricular tachycardia, the centers of rotation for the electrical activation correlated with their mechanical counterparts but do not exactly coincide \cite{Christoph2018,Ozsoy2021a,MolaviTabrizi2022}. The difference between electrical and mechanical activation patterns can be in part attributed to global coupling effects: a contraction in one part of the heart will deform it and thereby also stretch other parts. Hence, to accurately reconstruct electrical activation, methodologies are required to reconstruct the electrical signal from local deformation measurements.\\

To solve this demanding inversion task, classical inversion schemes like data assimilation\cite{Beam2020,Kovacheva2021,Lebert2019} have been presented, as well as supervised machine learning methods\cite{Christoph2020,Lebert2023}. The latter shows much potential as it can rely on an ever growing collection of newly developed machine learning and optimization techniques. They can, without much explicit knowledge of the specific system, accurately interpolate between seen data and compute these almost instantaneously, once trained. However, in the fully supervised methodology, training the large models can bring very high computational costs with them. As there are no diverse and complete experimental data sets available in this specific problem, there is also a need to create them through simulation, increasing the computational costs even more. Additionally, large supervised models are still seen as black boxes, while having difficulty in explaining certain outcomes or behaving well when applied outside their training domain. These problems are aggravated when data is sparse. Finally, it is practically difficult in a supervised learning framework to directly impose well-known physics laws on the solution or encode patient-specific information if available.\\

In this paper, we explore the use of an unsupervised machine-learning method for electromechanical inversion, known as physics-informed neural networks (PINN). PINNs have recently been developed to approximate solutions of partial differential equations\cite{Raissi2019}. On the one hand, PINNs use the expressiveness of neural networks and their applicability to large, non-linear and global optimization tasks. On the other hand, explicit physical or physiological knowledge is utilized by directly incorporating partial differential equations into the loss function. In addition, PINNs learn only one instance of a solution instead of the entire manifold, resulting in computationally fast optimizations and adaptable in- and outputs based on the specific situation. As such, diverse forward and inverse problems in a variety of applications have recently been approached using PINNs such as fluid dynamics\cite{Raissi2020a}, climate modeling\cite{Kashinath2021a} or transport\cite{Mo2021a}. Within cardiac electrophysiology context, PINNs have shown to be valuable in interpolating sparse intra-cardiac electrical activation times and estimating fiber directions\cite{SahliCostabal2020,Herrera2022} as well as characterizing tissue parameters of electrophysiological models in optical mapping data\cite{HerreroMartin2022}. PINNs have also already been used in cardiac mechanics problems, mainly to accelerate the evaluation of the forward model\cite{Dalton2023, Buoso2021}.\\ 

In this work we apply PINNs to recover the active mechanical tension field from local deformation measurements. As we aim to only offer proof-of-concept here, we use a simplified 2D geometry and linear elasticity laws, i.e. the Navier-Cauchy equations. Nonetheless, we demonstrate below that it is feasible to uncover excitation patterns even under very noisy conditions or heavily reduced spatial resolution. We also show how the PINN can easily be applied to different and challenging input data such as projected components on three 1D-planes. For extremely sparse or noisy conditions, additional regularization in the form of the eikonal equation, similar to previous work on electrical activation times\cite{SahliCostabal2020}, can be beneficial. In this extension we first optimize an initial network and calculate approximate activation times. These are then used as input for a second PINN optimization based on the eikonal equation. We also discuss the choices of hyperparameters such as network parameters and loss function weighting factors, as these are key in every optimization problem.

\section*{Methods}
\label{sec: methods}


\subsection*{Generation of synthetic data}
\label{subsec: data generation}

\subsubsection*{Forward model}
\label{subsubsec: physics}

The synthetic data set used in this work was generated by modeling electrical activation waves which propagate through a 2D elastic medium. The physical equations for the excitation wave, coupling and mechanical deformation as well as the parameter values follow the work of a previous publication\cite{Dierckx2015}. First, modified Aliev-Panfilov kinetics\cite{Aliev1996} are used to model the excitation wave dynamics:

\begin{gather}
    \frac{\partial v}{\partial t} =  \Delta v - kv(v-a)(v-1) - vw
    \label{A-P-1-eq}\\
    \frac{\partial w}{\partial t} = \epsilon(v)(kv-w)
    \label{A-P-2-eq}
\end{gather}

where v and w are the normalized transmembrane voltage and recovery variable, respectively. The step function $\epsilon(v)$ is equal to $0.1$ for $v\geq a$ and $1.0$ for $v<a$ as in previous work\cite{Pravdin2015}. Parameter values were set to $k=8$, $a=0.05$. This model has dimensionless space units (s.u.) and time units (t.u.). The action potential duration for $90\%$ repolarization ($APD_{90}$) was measured in simulations to be $6.8\,$t.u.. Scaling this to a human ventricular $APD_{90}$ of $270$\,ms\cite{Li1998,Morgan1992a}, 1 time unit in the model corresponds to $40$\,ms. Similarly, the measured conduction velocity for a one-dimensional isolated traveling pulse was $1.6$\,s.u./t.u. By comparing this to a tabulated value of wave propagation in human ventricle ($0.68$\,mm/ms\cite{Taggart2000a}), we conclude that one space unit in our model corresponds to $17$\,mm. In a next modeling step, the transmembrane voltage $v$ induces active tension $T_a$ in the medium according to the phenomenological ordinary differential equation\cite{Panfilov2005}:

\begin{equation}
    \frac{\partial T_a}{\partial t} = \epsilon(v)(k_Tv-T_a),
    \label{coupling-eq}
\end{equation}
resulting in a similar $T_a$ wave, slightly delayed in time with respect to the voltage. We used $k_T=1.5$. Lastly, for the deformation resulting from the active tension $T_a$, the linear and isotropic Navier-Cauchy equations for mechanical equilibrium in an elastic medium are used\cite{Slaughter2001}

\begin{equation}
    (\lambda+\mu)\vec{\nabla}(\vec{\nabla}\cdot\vec{U}) + \mu\Delta(\vec{U}) = - \vec{\nabla} T_a
    \label{nc-eq}
\end{equation}

Here, the passive elastic stresses, involving the material constants $\lambda$ and $\mu$ and spatial derivatives of the deformation vector $\vec{U}$, are balanced by the active tension gradient. We used the parameter values $\lambda=1.25$, $\mu=1$. Note that since we work with linear elasticity and dimensionless active tension here, only the ratio $\lambda / \mu$ is fixed by our parameter choice.\\ 

For simplicity, we use a 2D square domain with no-displacement boundaries. This choice corresponds to the experimental setting where a left-ventricular tissue wedge is subtended in a rigid square frame. We are aware that our model is inherently limited, since we adopt linear elasticity, isotropy of excitation and contraction, and a simplified geometry and boundary conditions. Nonetheless, the set of Eqs. \ref{A-P-1-eq}-\ref{nc-eq} provides a minimal model for the study of excitation-contraction coupling in the cardiac muscle.

\subsubsection*{Simulations}
\label{subsubsec: simulations}

All simulations were carried out on a homogeneous and isotropic 2D square domain of size 100\,s.u. for a time duration equal to $60$\,t.u.. Due to the absence of mechano-electrical feedback and the small displacement regime, the physical equations could be solved sequentially, allowing for great flexibility in simulation choices. The Aliev-Panfilov equations for the evolution of the transmembrane voltage (Eqs. \ref{A-P-1-eq},\ref{A-P-2-eq}) were solved using the \verb!ithildin! package\cite{kabus2023ithildin}, a finite difference time-stepper in C++ with MPI-parallellization, which was developed in our group and made publicly available. Simulations were executed on a 2D regular grid with a spatial resolution of 0.5, while integrating in time using forward Euler stepping at a temporal resolution of 0.05. Standard Neumann boundary conditions for the transmembrane voltage were applied to close the equations. We adopted 2 different initial conditions. One corresponded to 2 focal pulses at $t=0$ and $t=20$, initiated by inserting external voltage ($v=1$) in a circular region of radius $r=1$ with centers located at ($0,0$) and ($75,25$), respectively. The second simulation involved a stable rotating spiral initiated via a S1-S2 protocol. Every 1 t.u. the transmembrane voltage was stored for all spatial grid points. Active tension was developed according to Eq. \ref{coupling-eq}. The ODE was solved for every spatial point on a reduced grid (resolution 2.5 x 2.5 s.u.) with the use of python package scipy\cite{Virtanen2020} and its integrate module integrate.odeint. The resulting active tension values were stored every 2 t.u.. Finally, we made use of the publicly available finite-element package Fenicsx\cite{Logg2010} to solve the Navier-Cauchy equations (Eq. \ref{nc-eq}). The regular grid points of the active tension solution were used as vertices for the quadrilateral elements of the finite element mesh, while the function space for both the active tension and deformation vector were set to the standard $P_1$ linear Lagrange element. No-displacement boundary conditions were applied to all four sides. Final data consisted out of active tension and x- and y-components of the deformation vector on a $40$\,x\,$40$ spatial grid (resolution $2.5$\,s.u.) for $30$ time steps (resolution $2$\,t.u.).\\

\begin{figure}[ht!] \centering
	\includegraphics[width = 0.5\textwidth]{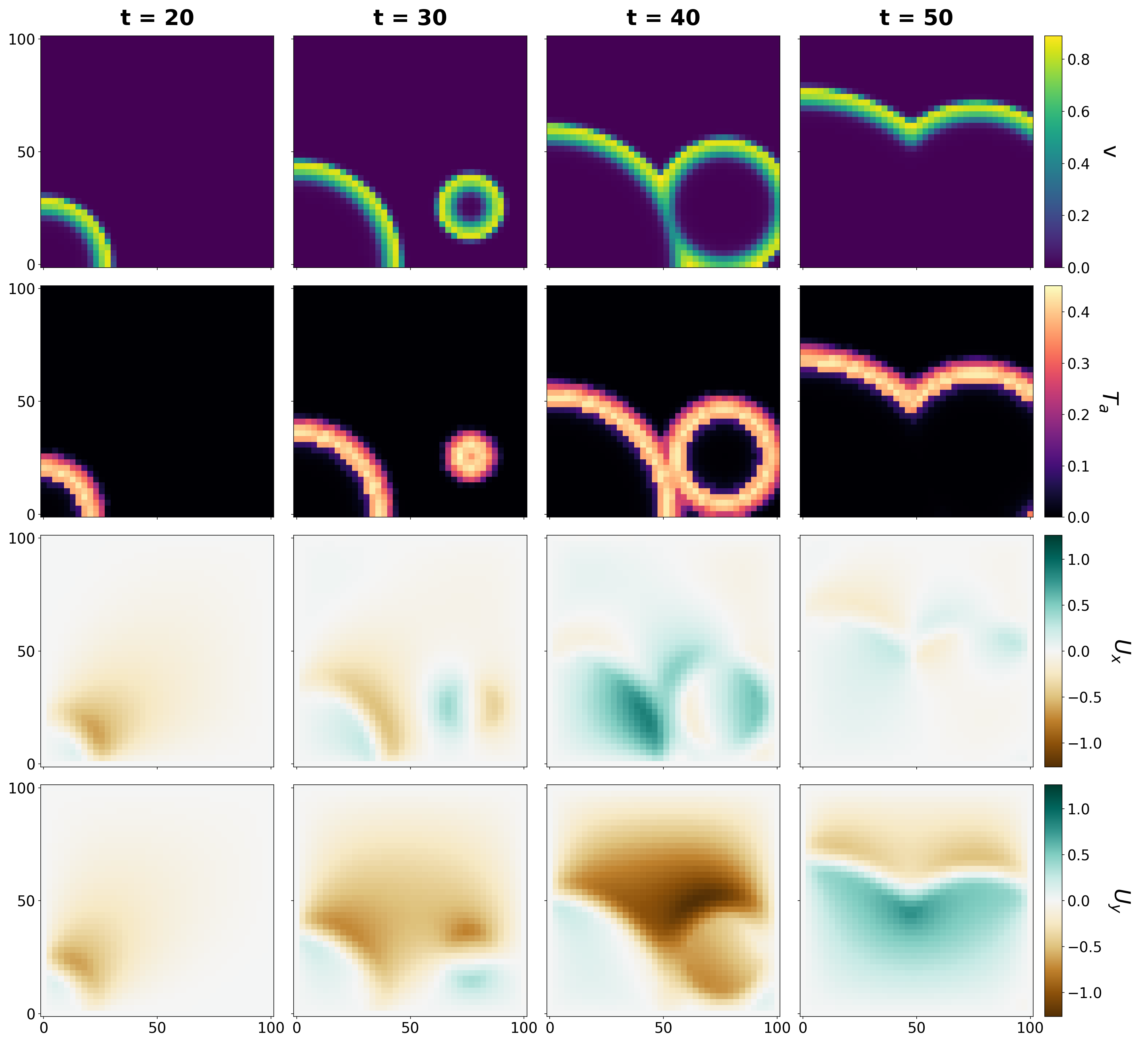}
	\caption{Time series of synthetic data generated as outlined in the Methods section. Two focal sources were initiated by inserting external voltage inside a circular region. The resulting transmembrane voltage wave produces a similar active tension wave, slightly delayed and broadened. The active tension then instantaneously induces deformation in the elastic medium.}
    \label{fig:focal_data}
\end{figure}

An example of a synthetic data set obtained by forward modeling is shown in Fig. \ref{fig:focal_data}. Two focal sources were initiated by stimulating a circular area of radius 1, differing in time and location. The active tension profile closely follows the voltage wave with only a small time delay and a slightly broader wave profile. As we are primarily focused in this study on recovering the accurate activation pattern, we will from here on out consider only the active tension $T_a$ as the variable of interest and investigate how it can be reconstructed from its resulting deformation. Both deformation components show large values in regions around the $T_a$-wave, while also exhibiting non-negligible values further away, due to the global spatial coupling in the Navier-Cauchy equations (Eq. \ref{nc-eq}).

\subsubsection*{Reduction of data quality}
\label{subsubsec: data reduction}

We tested the robustness of the developed method against the effect of noise, lower spatial resolution, and the absence of out-of-plane deformation components for tri-planar sampling. To this purpose, we processed the synthetic data in the following manner. We applied additive Gaussian noise to the simulated displacements $U_{x,0}$ and $U_{y,0}$. The noise level is taken proportional to the maximal absolute value of displacement in the x- and y-directions over the full spatiotemporal domain, i.e. $U_{x,max} = \mathrm{max}(U_{x,0})$, $U_{y,max} = \mathrm{max}(U_{y,0})$. Then, we have for every measurement point:

\begin{gather}
    U_{x} = U_{x,0} + \eta_{x}\ p\ U_{x,max} \\
    U_{y} = U_{y,0} + \eta_{y}\ p\ U_{y,max} 
\end{gather}

Here, all $\eta_{x}, \eta_{y}$ are sampled for every point in space and time from the standard normal distribution. The positive parameter $p$ is the noise level of choice and was varied between $0\%$ and $50\%$ in steps of $10\%$. Signal-to-noise ratios (SNR) were then calculated by dividing the variance of the original deformation values (signal) by the variance of the generated noise. This is done for both deformation components at every time step and afterwards averaged to obtain one SNR value for a specific data set. For a second group of data sets, we reduced the spatial resolution using a moving spatial average, since spatial resolution in an actual ultrasound recording depends on the scan sequence and hardware. Deformation components were averaged over $n^2$ data points in a $n$ x $n$ square resulting in images used for inversion which have $40/n$ by $40/n$ pixels. Our third type of non-ideality was inspired by tri-planar acquisition\cite{Ramalli2020}, in which three planes are recorded through the chambers, at an angle of $60 \degree$ to each other. Thus, three lines of each 40 points were sampled from the total data set, divided by $60 \degree$. As a true recording only contains information about the displacement within the viewing plane, the deformation vector $\vec{U}$ was projected onto that line, resulting in one projected deformation scalar per point on the line, yielding only 3$\times 40 = 120$ data points per time frame to be used as input data for the inversion. 

\subsubsection*{Quality assessment of the reconstruction}
\label{subsubsec: quality assesment}

To evaluate the quality of reconstruction, we did not adapt a pixel-based metric such as the root-mean square error or the L2-difference between the ground truth and the reconstructed image. The underlying reason is that, in view of localizing the electrical sources of arrhythmia, we are primarily interested in the accurate shape of the activation wave rather than the absolute values. Additionally, there is only information present about the gradient of the active tension field through the Navier-Cauchy equations, such that the absolute baseline value of $T_a$ cannot be determined from deformation alone.  Therefore, we adopted in this work a well-established image quality measure called the Feature Similarity (fsim) index\cite{Zhang2011}. The fsim index is calculated from the phase congruence, a dimensionless quantity of the significance of a local structure, as well as the image gradient magnitude, see details in the original paper \cite{Zhang2011}. By this choice, we will have no undesired effects from higher background values and our results are evaluated with the focus on the global shape. Fsim values shown in this work for time series are calculated separately for every frame and then averaged over the time series to obtain a single score between 0 and 1 that covers the quality of reconstruction. 

\subsection*{Inversion methods}
\label{subsec: models}

Our working hypothesis is that the inversion of sparse ultrasound measurements can be translated to a large optimization problem, where the unknown is the full spatiotemporal active tension field. Such large-scale optimization problem can be tackled by physics-informed neural networks, in which the neural networks that were previously designed for supervised machine learning tasks, are used as flexible non-linear function approximators. They are equipped with efficient optimization schemes and minimize loss functions which incorporate explicit physical knowledge, thus combining data-driven and model-driven methodologies.

\subsubsection*{Navier-Cauchy Physics-Informed Neural Network (NC-PINN)}
\label{subsubsec: NC PINN}

\begin{figure}[ht!] \centering
	\includegraphics[width = 0.75\textwidth]{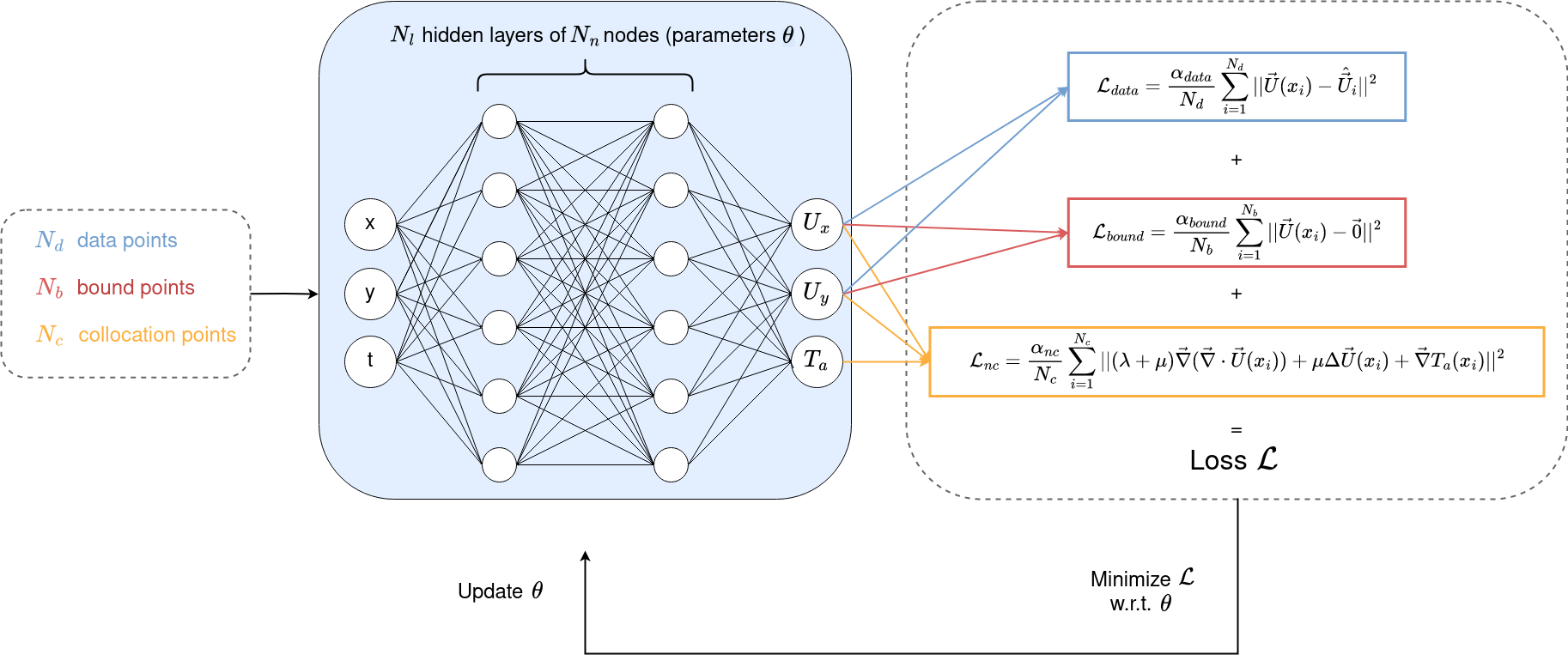}
	\caption{Schematic overview of one iteration of the optimization process in the NC-PINN. The neural network consist of a densely connected architecture of $N_l$ hidden layers with $N_n$ nodes each. Inputs are spatiotemporal coordinates $(x,y,t)$, outputs are deformation components $U_x$, $U_y$ and active tension $T_a$. Colours indicate the points (coordinates) used for each loss term. The network is updated after the total loss is minimized with respect to the network parameters $\theta$ (weights and biases of the network).}
    \label{fig:standard-pinn}
\end{figure}

To reconstruct waves of active tension from deformation data, a physics-informed neural network was implemented in TensorFlow\cite{tensorflow2015-whitepaper} (version 2.10.0), based on the point-wise formulation with a fully-connected, dense structure\cite{Raissi2019} ($N_l$ hidden layers of each $N_n$ nodes), see Fig. \ref{fig:standard-pinn}. The neural network is used as a flexible function approximator, which maps spatial and temporal coordinates ($x,y,t$) to both deformation components ($U_x, U_y $) and active tension ($T_a$). We kept the $\vec{U}$- and $T_a$-prediction in the same network (e.g. in contrast to other work\cite{SahliCostabal2020}), such that the identified spatiotemporal patterns in the deformation space can easily be shared with the active tension prediction. PINN optimization involves the minimization of a custom loss function that will ensure a match with the measured data, obey physical laws via differential equations and respect certain boundary conditions. The relative importance of these contributions is tuned via weighing factors $\alpha_*$, to be discussed below. The following loss function $\mathcal{L}$ was chosen in this work: 

\begin{gather}
    \mathcal{L} = \mathcal{L}_{data} + \mathcal{L}_{nc} + \mathcal{L}_{bound}
    \label{total-loss-eq}\\
    \mathcal{L}_{data} = \frac{\alpha_{data}}{N_d}\sum\limits_{i=1}^{N_d} || \vec{U}(x_i) - \hat{\vec{U}}_i||^2
    \label{loss-data-eq}\\
    \mathcal{L}_{nc} = \frac{\alpha_{nc}}{N_c}\sum\limits_{i=1}^{N_c} || (\lambda+\mu)\vec{\nabla}(\vec{\nabla}\cdot\vec{U}(x_i)) + \mu\Delta \vec{U}(x_i) + \vec{\nabla} T_a(x_i)||^2
    \label{loss-nc-eq}\\  
    \mathcal{L}_{bound} = \frac{\alpha_{bound} }{N_b}\sum\limits_{i=1}^{N_b} || \vec{U}(x_i) - \vec{0} || ^2
    \label{loss-bound-eq}
\end{gather}

The first term $\mathcal{L}_{data}$ calculates the mean squared error between the predicted deformation components $\vec{U}(x_i)$ and the deformation data that is available $\hat{\vec{U}}_i$. This is done for $N_d$ randomly sampled data points. Dividing by the number of points converts the sum to averages, such that the value is independent of the chosen number of sampled points. The second term $\mathcal{L}_{nc}$ imposes that the output satisfies, at least locally, the Navier-Cauchy equations and is thus formulated in terms of spatial derivatives of the deformation and active tension as well as the mechanical parameters of the equation, $\lambda$ and $\mu$. Both parameters are assumed to be known in this study. Derivatives are conveniently calculated through automatic differentiation in the same manner as gradients for the NN parameters $\theta$ are obtained. Because the physical equations are supposed to hold in the whole spatiotemporal domain, $N_c$ collocation points are sampled using the Latin hypercube design. Lastly, the term $\mathcal{L}_{bound}$ translates the no-displacement boundary conditions. Similarly, $N_b$ collocation points are sampled through the Latin hypercube design, while setting here either the x- or y-coordinate equal to $0$ or L with L the size of the spatial domain in order to obtain boundary points.\\

Once the total loss function $\mathcal{L}$ is calculated, network parameters $\theta$ can be updated based on the gradients of the minimization of $\mathcal{L}$ with respect to $\theta$. This whole process is done for $N$ iterations, also called mini-batches. We have set $N=10^{4}$ and chose $64$ points per mini-batch meaning $N_d=N_c=N_b=64$. The optimization is done with the Adam optimizer of TensorFlow and its default parameters. A schematic overview of one iteration is shown in Fig. \ref{fig:standard-pinn}. To facilitate the optimization, spatial and temporal coordinates were first normalized to the range $\left[0,1\right]$, while the active tension output was put through the TensorFlow sigmoid function and multiplied by a maximum value before calculating the loss function. This approach is possible as $T_a$ will always be non-negative and an upper limit can be well estimated, similar to the magnitude of velocity in other studies\cite{SahliCostabal2020}. Lastly, the hyperbolic tangent function was used as the differentiable activation function and Glorot initialization from a uniform distribution has been applied to all NN parameters before the start of optimization.

\subsubsection*{Additional regularization using wave propagation constraint (EIK-PINN)}
\label{subsubsec: wex}

In the case of poor data quality, we propose to use additional physical information. For simple wave patterns without repeated activity, the isotropic and homogeneous eikonal equation \cite{Franzone1993} can be used. This viewpoint relates the local activation time $\tau(\vec{r})$ to the local conduction velocity $v(\vec{r})$:

\begin{equation}
    \lVert \vec{\nabla}\tau(\vec{r}) \rVert = \frac{1}{v(\vec{r})}.
    \label{eik-eq}
\end{equation}

Although detailed measurements may allow to estimate a local conduction velocity\cite{SahliCostabal2020}, $v(\vec{r})$ was taken to be constant in space in this work. Moreover, curvature effects on the propagation speed\cite{Kuramoto1984,Dierckx2011} are neglected. A schematic overview of the method combining the elastic and wave propagation regularization is provided in Fig. \ref{fig:slices}b. First, the deformation data was fed into the NC-PINN as outlined in Fig. \ref{fig:standard-pinn}, to reconstruct a first estimate of the active tension, denoted $\hat{T}_a$. For every spatial coordinate on a fine regular grid, the initial local activation time $\hat{\tau}$ was then found by searching for the time when $\hat{T}_a(t)$ attains its maximum over the full time domain. Only points for which $T_a(\hat{\tau}) \geq  T_{a,max}/3$, with $T_{a,max}$ a predefined constant, were kept. Finally, these $\hat{\tau}$ values were used as input to a second PINN, with wave propagation regularization, see Fig. \ref{fig:slices}a. Since this PINN serves as an approximator to the function $\tau(\vec{r})$, it only takes $x$ and $y$ as input, not time. The loss function for this network was: 

\begin{gather}
    \mathcal{L}' = \mathcal{L}_{data-\tau} + \mathcal{L}_{eik}
    \label{NN2-loss-eq}\\
    \mathcal{L}_{data-\tau} = \frac{\alpha_{data-\tau}}{N_d}\sum\limits_{i}^{N_d} (\tau(x_i) - \hat{\tau}_i)^2
    \label{loss-data-tau-eq}\\
    \mathcal{L}_{eik} = \frac{\alpha_{eik}}{N_c}\sum\limits_{i}^{N_c} \left(\lVert \vec{\nabla}\tau(x_i) \rVert-\frac{1}{v}\right)^2. 
    \label{loss-eik-eq}
\end{gather}

The data term consists of the pre-calculated $\hat{\tau}$ values, while the eikonal term supplements additional information in order to inter- or extrapolate activation times or to overwrite potential errors from the previous NC-solution. To return to the $T_a(x,y,t)$ space, we make use of a predefined $T_a(t)$ shape to approximate the time evolution of typical active tension progression. In this work we have assumed that the tension developed over time follows a Gaussian curve, a simplifying but sufficient function, of given width $\sigma$: 
\begin{equation}
    T_a(t;\tau) = T_{a,max} \exp\left(\frac{(t-\tau)^2}{2\sigma^2} \right)
    \label{gaussian-eq}
\end{equation}

with constant parameters $T_{a,max}$ and $\sigma$. We have chosen $v=1.5$, $\sigma$ = $2.5$ and $T_{a,max} = 0.42$, as approximate values of the measurements from the data set ($v_{exp}=1.6$, $\sigma_{exp} = 2.3 $, $T_{a,max,exp} = 0.42$). In applications, these values will have to be either estimated based on patient-specific data, on literature values, or in the case of the conduction velocity, could be predicted simultaneously with the activation times.

\begin{figure}[ht!]
    \centering

     \begin{subfigure}{0.6\textwidth}
        \includegraphics[width=\linewidth]{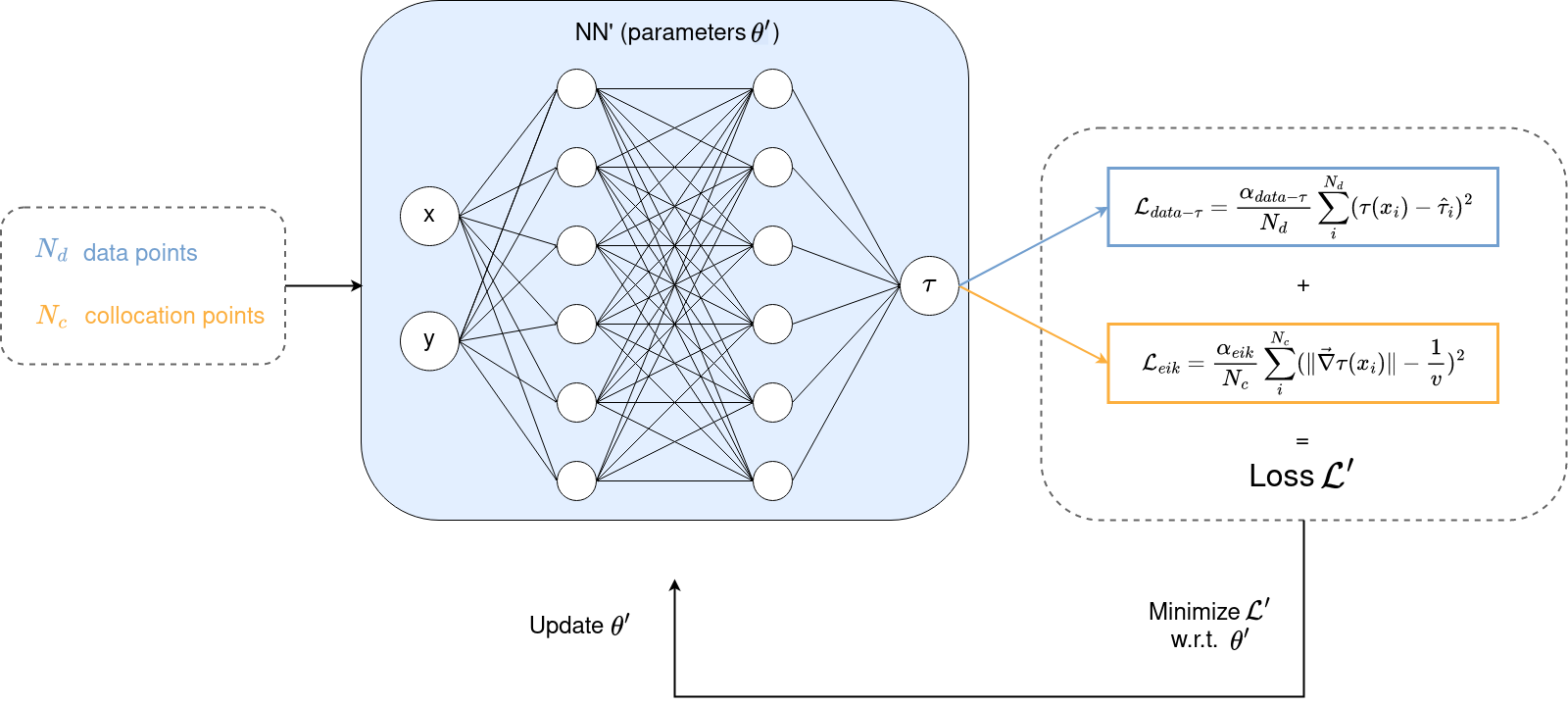}
        \caption{}
        \label{fig:EIK-pinn-optimization}
    \end{subfigure}

    \vspace{1cm}
    
    \begin{subfigure}{0.7\textwidth}
        \includegraphics[width=\linewidth]{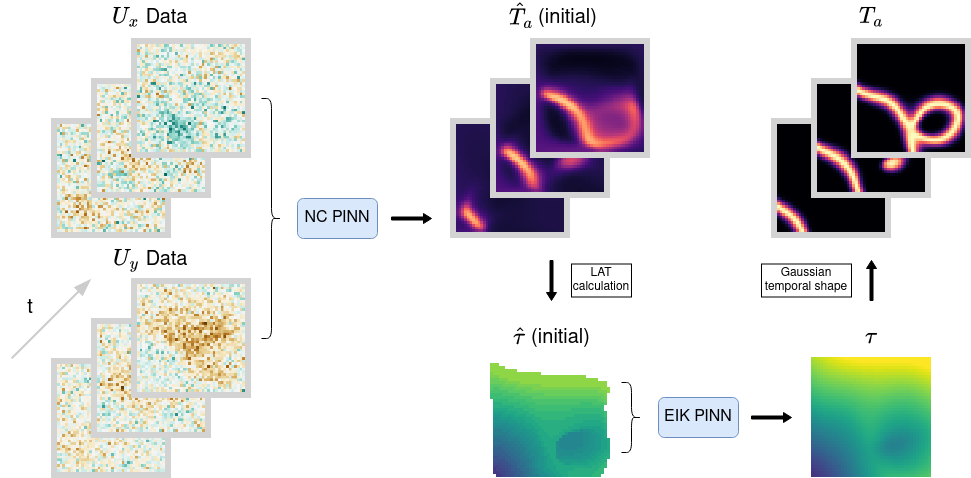}
        \caption{}
        \label{fig:EIK-pinn-scheme}
    \end{subfigure}
    
    \caption{\textbf{(a)} Overview of the EIK-PINN to enhance the reconstruction of the local activation time (LAT) $\tau$. Colours indicate the points (coordinates) used for each loss term.
    \textbf{(b)} Schematic overview of the two-step optimization (NC+EIK) in case both elastic and wave propagation constraints are used. The standard NC-PINN is used to convert deformation $\vec{U}$ into a first estimate of active tension $\hat{T}_a$. From this field, the LAT is calculated, and optimized subject to the eikonal relation (Eq. \ref{eik-eq}) via a second PINN, depicted in panel (a). The emerging $\tau$ can either inter- or extrapolate the solution or overwrite initially difficult regions, after which it is converted back to $T_a$ via Eq. \ref{gaussian-eq}.}
    \label{fig:EIK-PINN}
\end{figure}

\section*{Results}
\label{sec: results}

\subsection*{Inversion using the NC-PINN}
\label{subsec:standard}

\begin{figure}[ht!]
    \centering
    
    \begin{subfigure}{0.49\textwidth}
        \includegraphics[width=\linewidth]{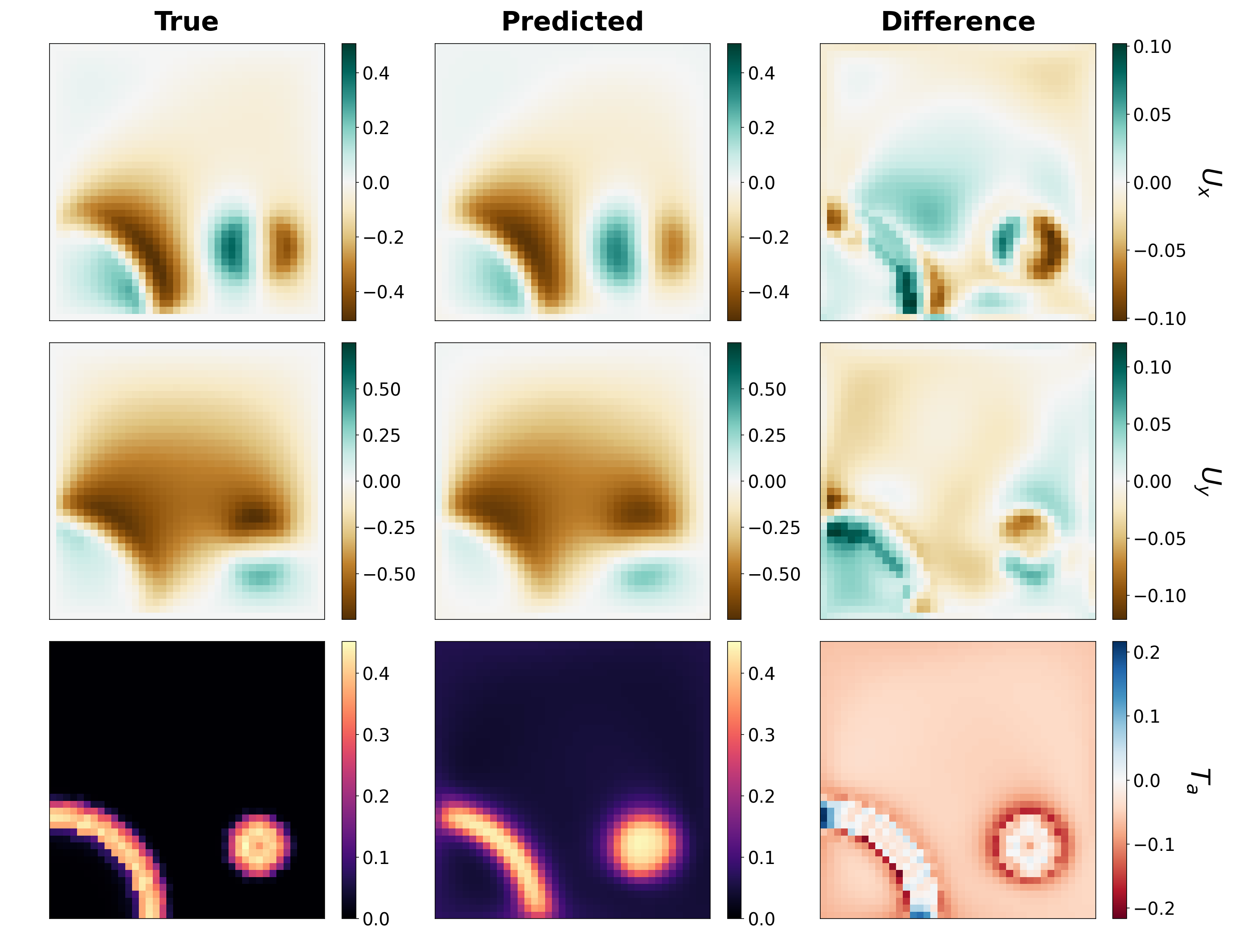}
        \caption{}
        \label{fig:basic_results-a}
    \end{subfigure}
    \hfill
    \begin{subfigure}{0.49\textwidth}
        \includegraphics[width=\linewidth]{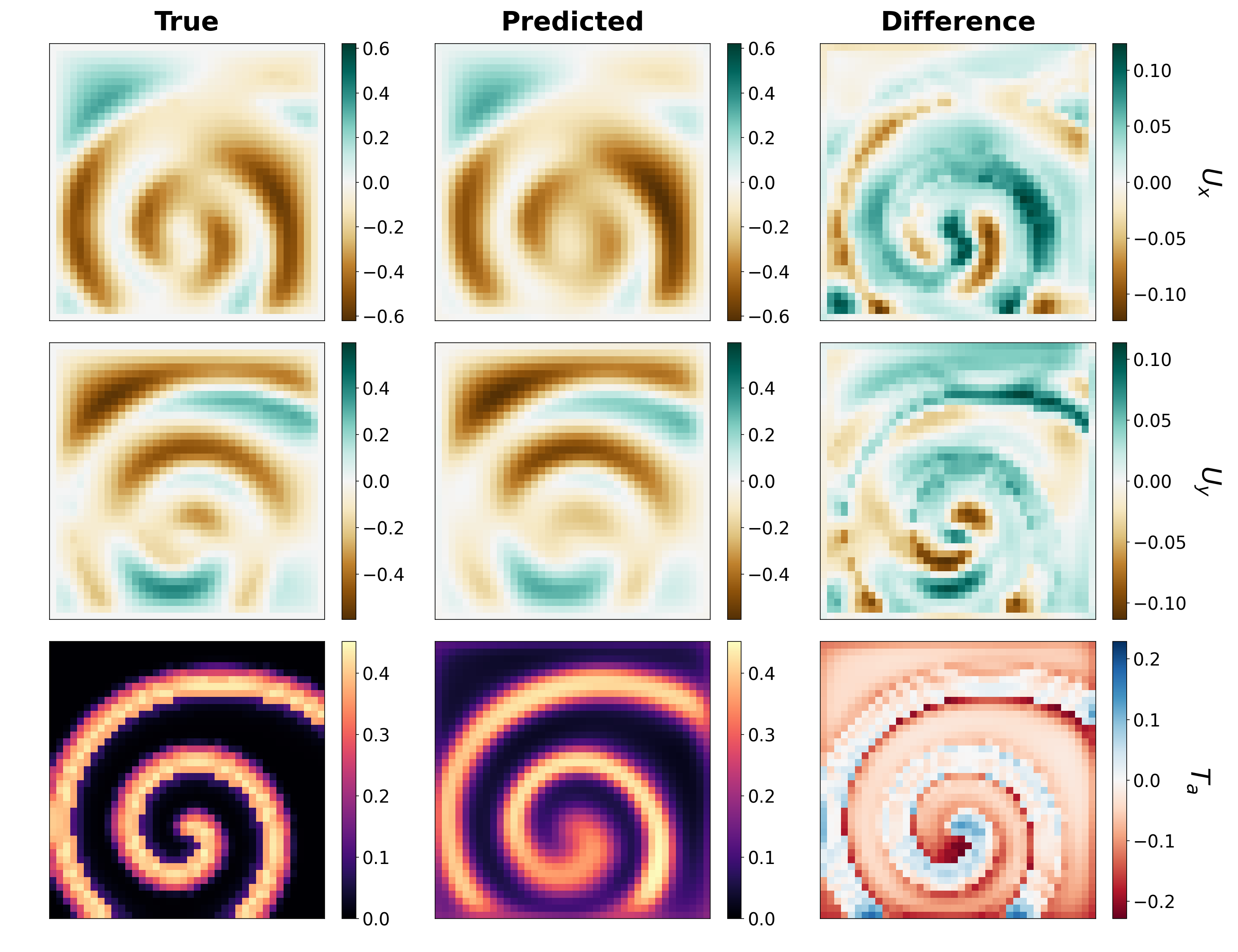}
        \caption{}
        \label{fig:basic_results-b}
    \end{subfigure}
    
    \begin{subfigure}{0.49\textwidth}
        \includegraphics[width=\linewidth]{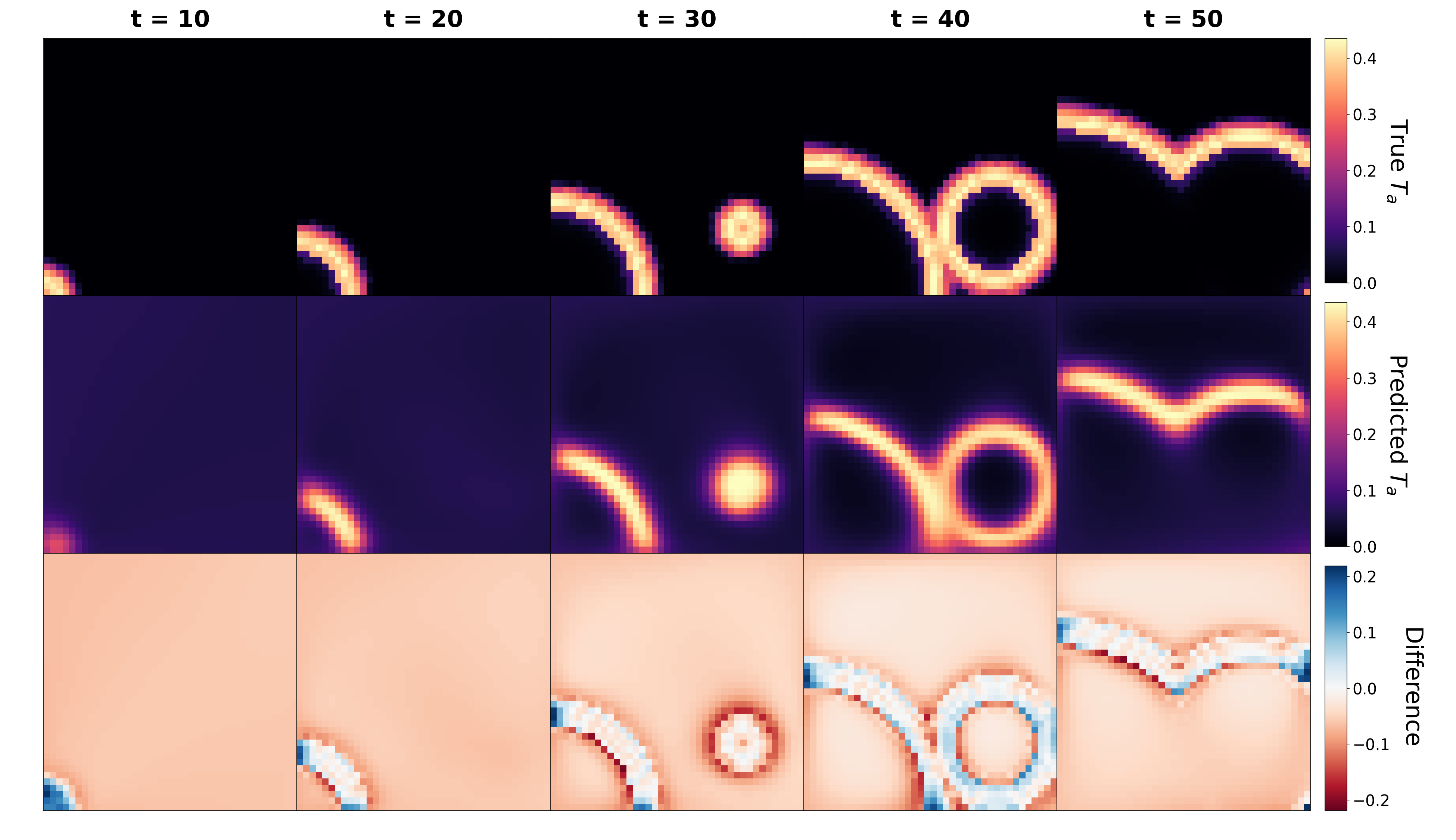}
        \caption{}
        \label{fig:basic_results-c}
    \end{subfigure}
    \hfill
    \begin{subfigure}{0.49\textwidth}
        \includegraphics[width=\linewidth]{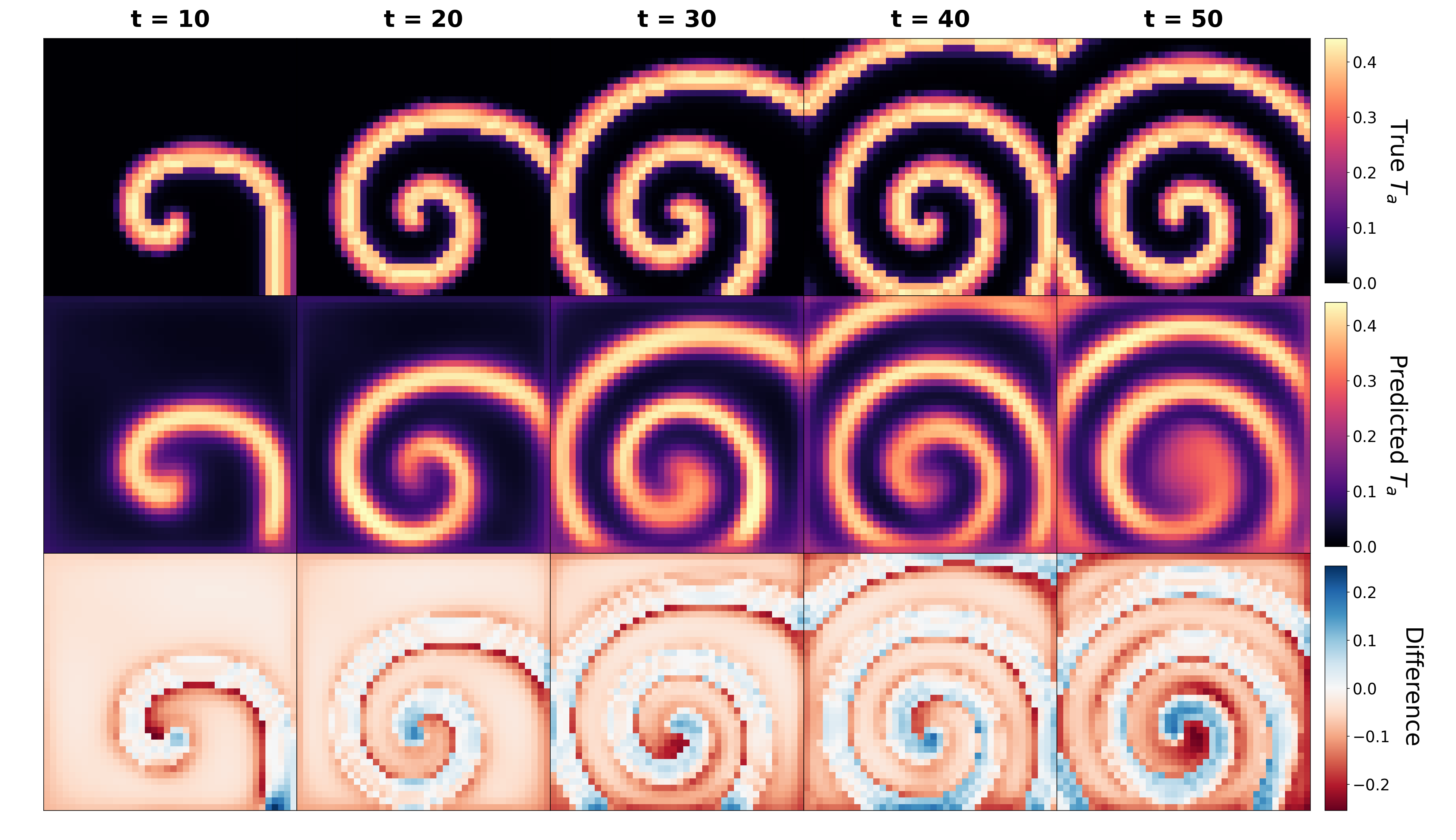}
        \caption{}
        \label{fig:basic_results-d}
    \end{subfigure}
    
    \caption{NC-PINN results for the reconstruction of an active tension wave from mechanical displacement data using the optimization hyperparameters in Table \ref{tab: hyperparameters}. Panels \textbf{(a)} and \textbf{(b)} show the true and predicted spatial fields $U_x$, $U_y$ and $T_a$ as well as their differences, at time $t=30$. Panels \textbf{(c)} and \textbf{(d)} visualise true and predicted snapshots of $T_a$ over time and their difference.}
    \label{fig:basic_results}

\end{figure}

First, we performed PINN optimization on all available data, i.e. both deformation components were densely sampled in space and time without any noise. The hyperparameters of the NN and loss function are summarized in Table \ref{tab: hyperparameters} and their influence will be discussed below. The optimization was executed on a HP Z-book, 11th Gen Intel Core i7-11800H, while TensorFlow's automatic CPU parallelization used all 16 available virtual nodes. Computational times were measured to be $196.11\pm3.08$\,s for the focal pattern, increasing slightly to $204.14\pm2.56$\,s for the spiral (10 independent runs). Fig. \ref{fig:basic_results} shows typical reconstructions of the deformation and active tension, for focal activity in panel (a) and spiral-shaped activation in panel (b). The deformation fields are accurately fitted spatially, and the point-wise relative error remains smaller than $20\%$ of the maximal attained value. The largest deviations are primarily located in regions where sharp gradients are present. Such behaviour is expected, since these regions are naturally the most challenging to represent via a continuous function approximator. Panels (c) and (d) show snapshots in time of the reconstructed active tension. Overall, the pattern is well represented over the whole time-domain for both the focal and spiral data sets. The following minor discrepancies are present in the reconstruction. The asymmetric property of the wave shape, i.e. a sharper upstroke and broader repolarization process, is not captured by the reconstruction. Second, the no-displacement boundary condition results in very small regions at the border where the full active tension magnitude is reduced. Lastly, the center of the spiral pattern becomes more diffuse at later time steps. Nevertheless, despite the global spatial coupling, the reconstruction of the cause (i.e. local tension) from the effect (i.e. deformation) seems to be feasible in this setting. 

\begin{table}[ht!]
    \centering
    \resizebox{0.5\columnwidth}{!}{
    \begin{tabular}{|c|c|c|c|c|c|c|c|}
    \hline
    \multicolumn{1}{|c|}{\textbf{Pattern}} & \multicolumn{1}{c|}{\textbf{N}} & \multicolumn{1}{c|}{$\mathbf{N_d}$, $\mathbf{N_c}$, $\mathbf{N_b}$
    } & \multicolumn{1}{c|}{$\mathbf{N_l}$} & \multicolumn{1}{c|}{$\mathbf{N_n}$} & \multicolumn{1}{c|}{$\mathbf{\alpha_{data}}$}  & \multicolumn{1}{c|}{$\mathbf{\alpha_{bound}}$}  & \multicolumn{1}{c|}{$\mathbf{\alpha_{nc}}$}\\
    \hline
    \hline
    Focal & $10^4$ & $64$ & $15$ & $35$ & $1$ & $1$ & $100$ \\
    \hline
    Spiral & $10^4$ & $64$ & $15$ & $50$ & $1$ & $1$ & $100$ \\
    \hline
    
    \end{tabular}}
    
    \caption{\label{tab: hyperparameters}Table of all hyperparameters used in the optimization of the NC-PINN.}
\end{table}

\begin{figure}[ht!]
    \centering
    \begin{subfigure}{0.35\textwidth}
        \includegraphics[width=\linewidth]{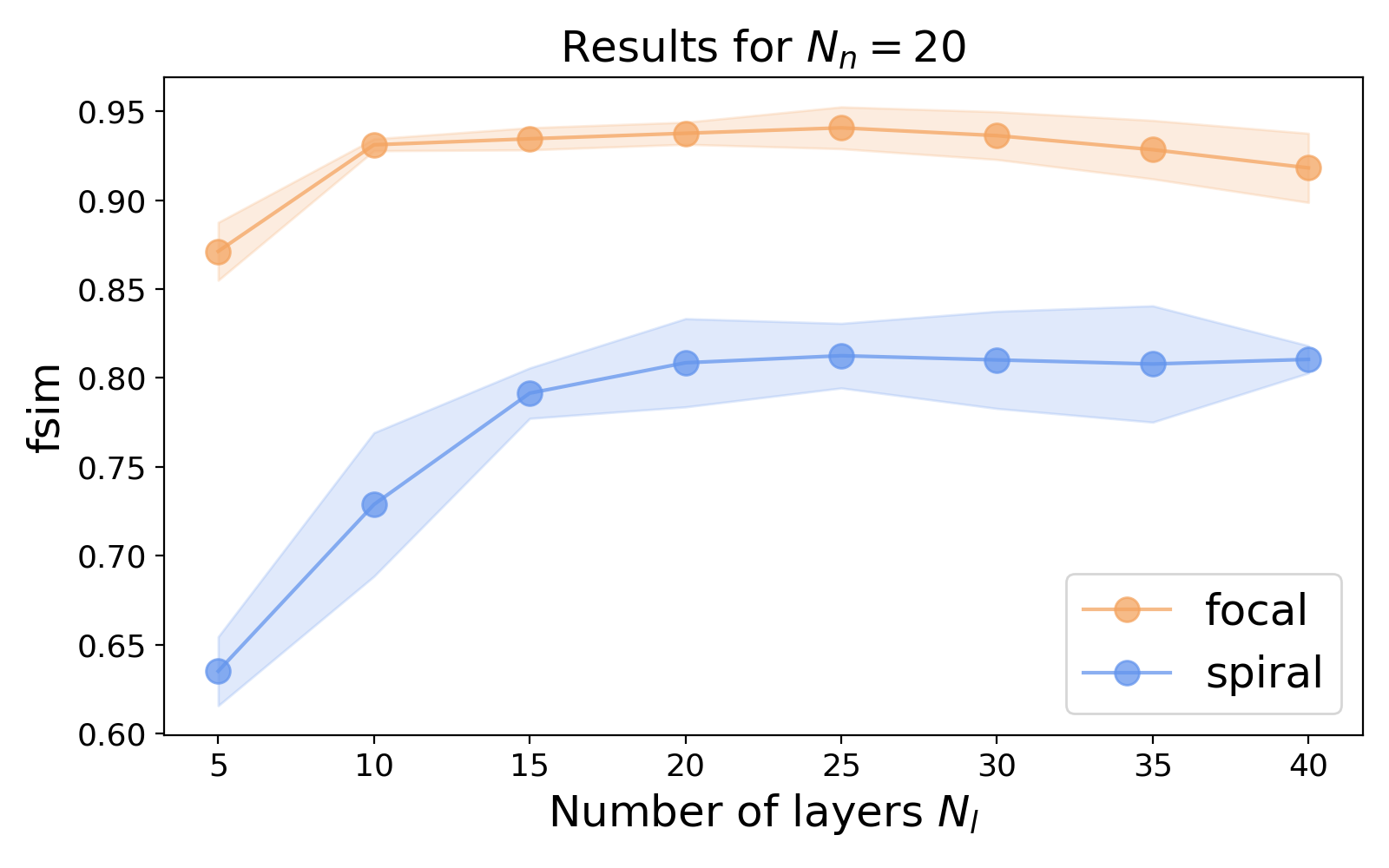}
        \centering
        \caption{}
        \label{fig: hyperparameters-a}
    \end{subfigure}
    \begin{subfigure}{0.35\textwidth}
        \includegraphics[width=\linewidth]{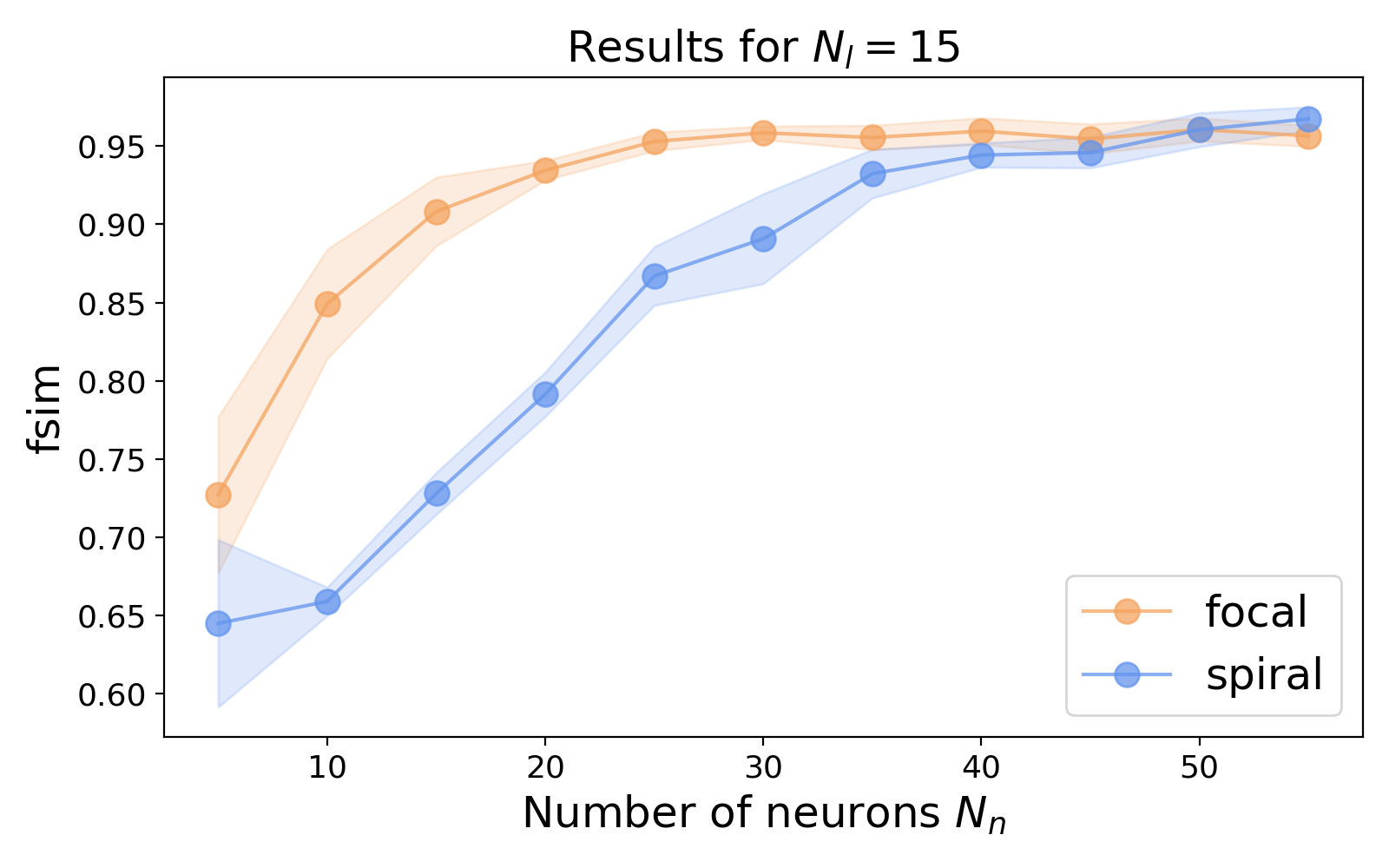}
        \centering
        \caption{}
        \label{fig: hyperparameters-b}
    \end{subfigure}
    \hspace{0.1\textwidth}
    \begin{subfigure}{0.35\textwidth}
        \includegraphics[width=\linewidth]{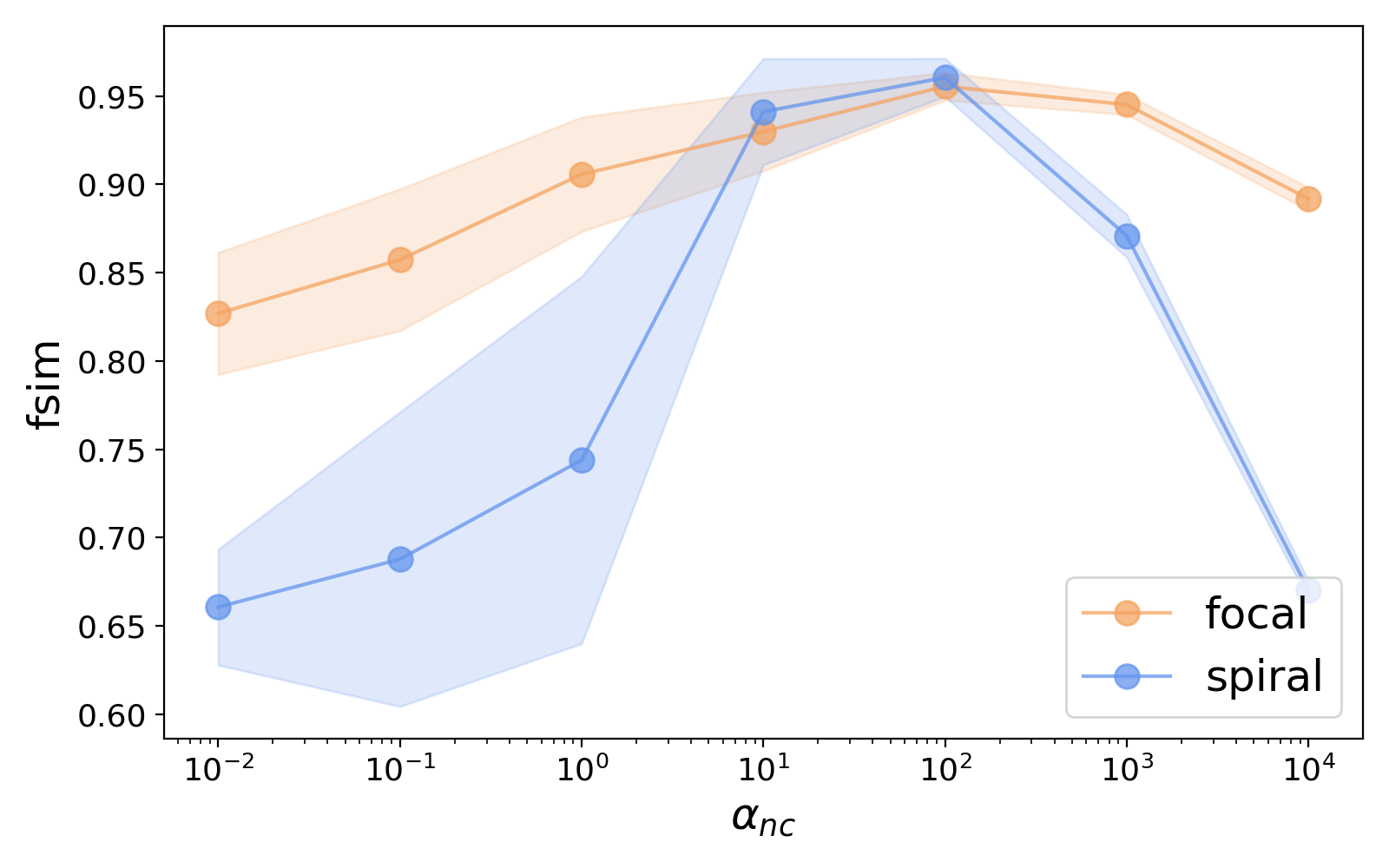}
        \caption{}
        \label{fig: hyperparameters-c}
    \end{subfigure}
    
    \caption{Effects of varying hyperparameters in optimizing the NC-PINN on all available data. Reported fsim values are averaged over $5$ independent runs, while the shaded area represents the standard deviation. Panels \textbf{(a)} and \textbf{(b)} show the influence on the fsim score for the number of layers $N_l$ and number of neurons $N_n$. All curves seem to reach a plateau phase, while their converging point depends on the pattern type. Panel \textbf{(c)} shows the influence of the weighting parameter $\alpha_{nc}$, while $\alpha_{data} = 1$. Both curves indicate an optimal value around $10^2$.}
    \label{fig: hyperparameters}

\end{figure}

To select the optimal hyperparameters such as number of nodes $N_n$, number of layers $N_l$ and the weighting factor for the Navier-Cauchy term $\alpha_{nc}$, we looked at how the time averaged fsim score varied across these parameters. Fig. \ref{fig: hyperparameters} presents the averages and standard deviations over five runs for both the focal and spiral pattern. To see the effect of NN choices i.e. $N_n$ and $N_l$, we kept $\alpha_{data} = 1$ and $\alpha_{nc} = 10^2$ constant. In Fig. \ref{fig: hyperparameters}A, the quality of reconstruction reaches a plateau quite fast and increasing the number of layers does not significantly improve the reconstruction. This happens after 10 layers for the focal and 15 layers for the spiral pattern. Fig. \ref{fig: hyperparameters}B investigates how many neurons per layer are needed for convergence. While the focal curve already flattens around 25 neurons, the spiral curve keeps significantly increasing until 40. Such behaviour can be expected, since more degrees of freedom are needed in the function approximation to represent a more complex wave profile. Larger networks demand more iterations, memory and are prone to potential overfitting, such that the network size should ideally be chosen close to the convergence point. Note, however, that overfitting is less of a concern in PINNs as the regularization via partial differential equations will reinforce continuous functions in a natural way, especially when dealing with accurate sparse data or Gaussian noise. As a consequence, the main objective of choosing $N_l$ and $N_n$ should be the minimally necessary capacity to express the expected spatiotemporal pattern. Based on the curves in Fig. \ref{fig: hyperparameters}A-B, we set $N_l = 15$ for both patterns, while we reconstructed the focal pattern with $N_n = 30$ and the spiral pattern with $N_n = 50$. Fig. \ref{fig: hyperparameters}C shows the quality of reconstruction for different strengths of physics-based regularization, by varying the weight $\alpha_{nc}$ of the Navier-Cauchy loss term. All these reconstructions have a constant $\alpha_{data}=1$, $N_l = 15$, $N_n = 35$ or $N_n = 50$. Both curves seem to have an optimal value around $\alpha_{nc} = 10^2$, indicating the optimal trade-off between data and physics-based loss terms. We will further examine the choice of weighting parameters in the Discussion section. Unless explicitly stated, all PINN optimizations in the remainder of this work will have the same hyperparameters summarized in Table \ref{tab: hyperparameters}.

\subsubsection*{Effect of noise}
\label{subsubsec: noise}

\begin{figure}[ht!]
    \centering
    \begin{subfigure}{0.35\textwidth}
    \centering
        \includegraphics[width=\linewidth]{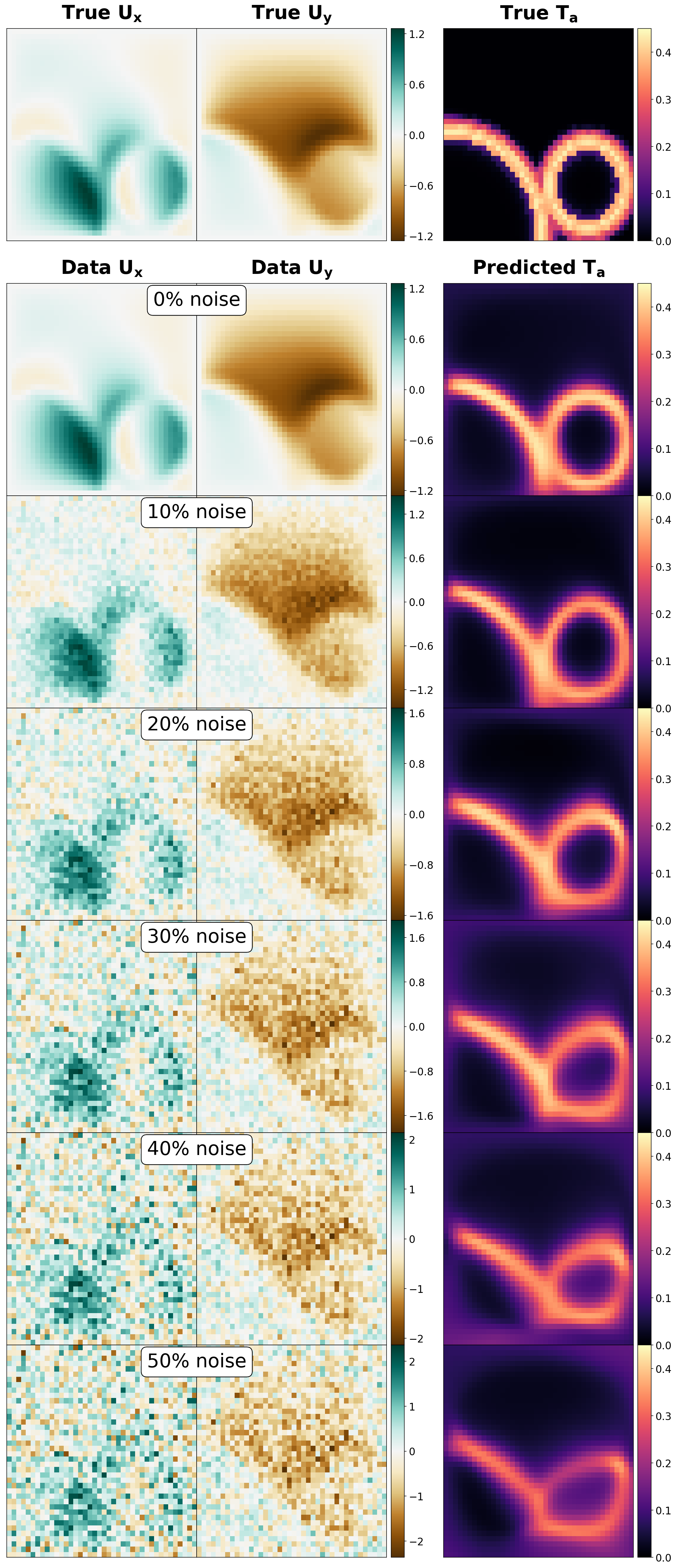}
        \caption{}
        \label{fig:noise1}
    \end{subfigure}
    \hspace{0.05\textwidth} 
    \begin{subfigure}{0.35\textwidth}
    \centering
        \includegraphics[width=\linewidth]{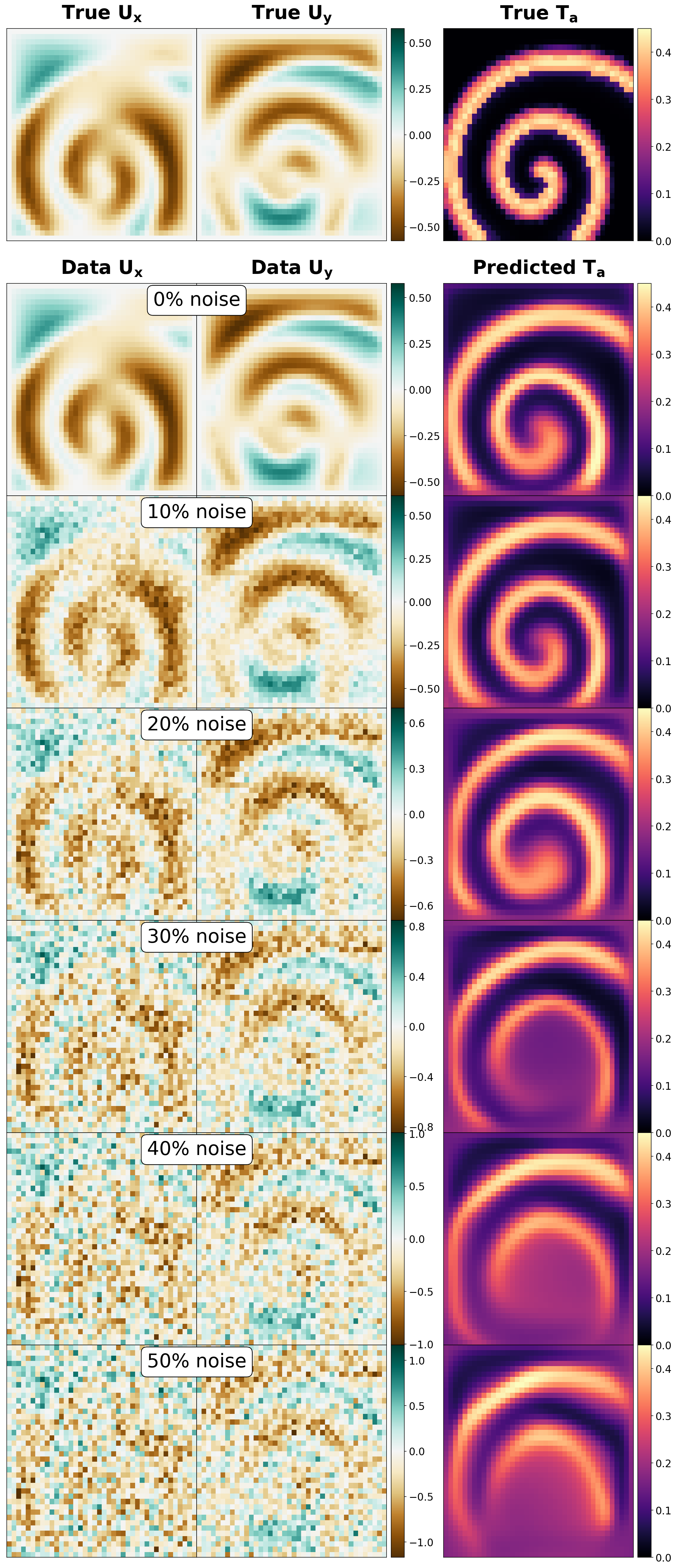}
        \caption{}
        \label{fig:noise2}
    \end{subfigure}
    \begin{subfigure}{0.35\textwidth}
    \hspace{-0.1\textwidth}
    \centering
        \includegraphics[width=\linewidth]{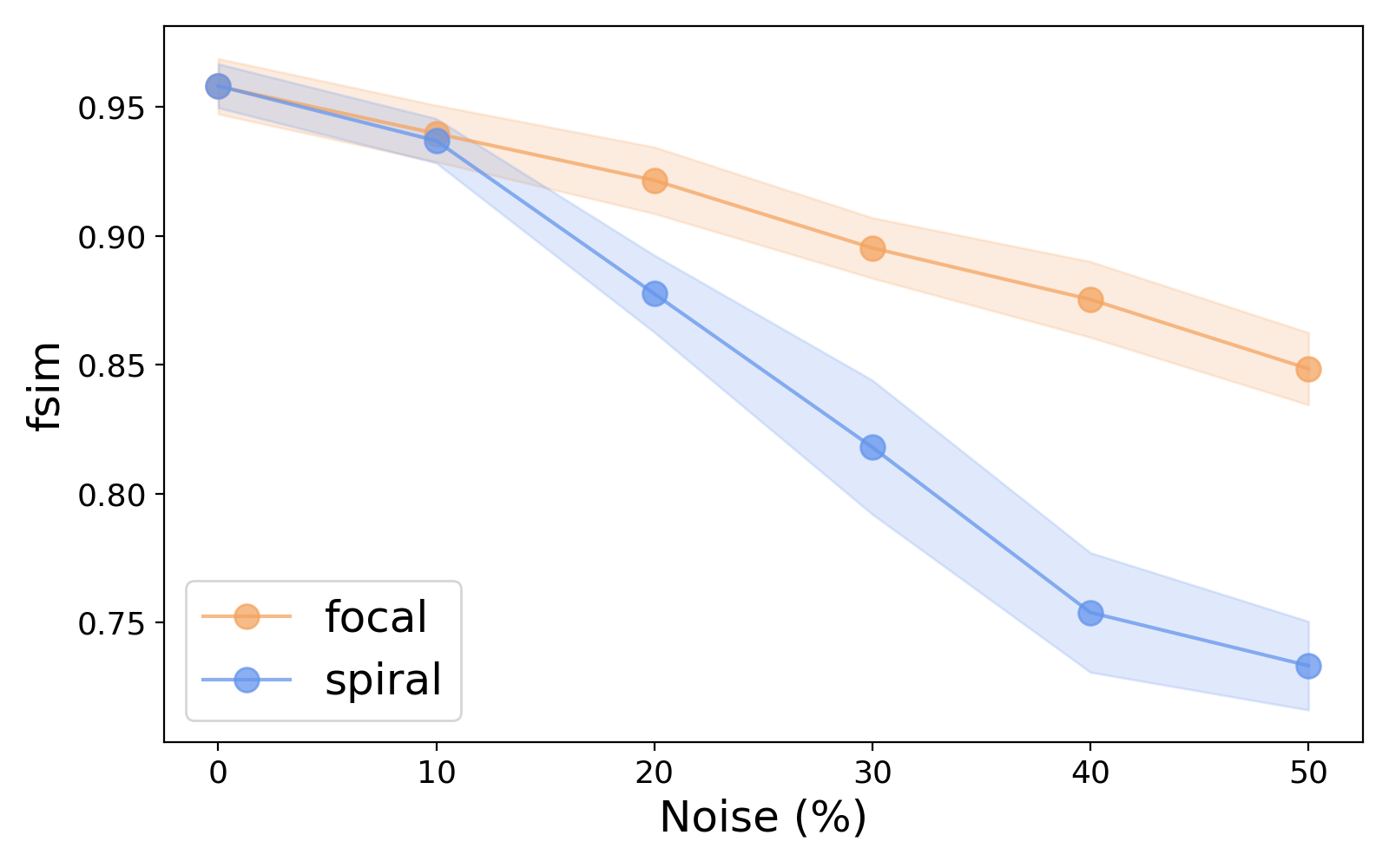}
        \caption{}
        \label{fig:noise3}
    \end{subfigure} 
    \hspace{0.05\textwidth}
    \begin{subfigure}{0.35\textwidth}
    \hspace{-0.1\textwidth}
    \centering
        \includegraphics[width=\linewidth]{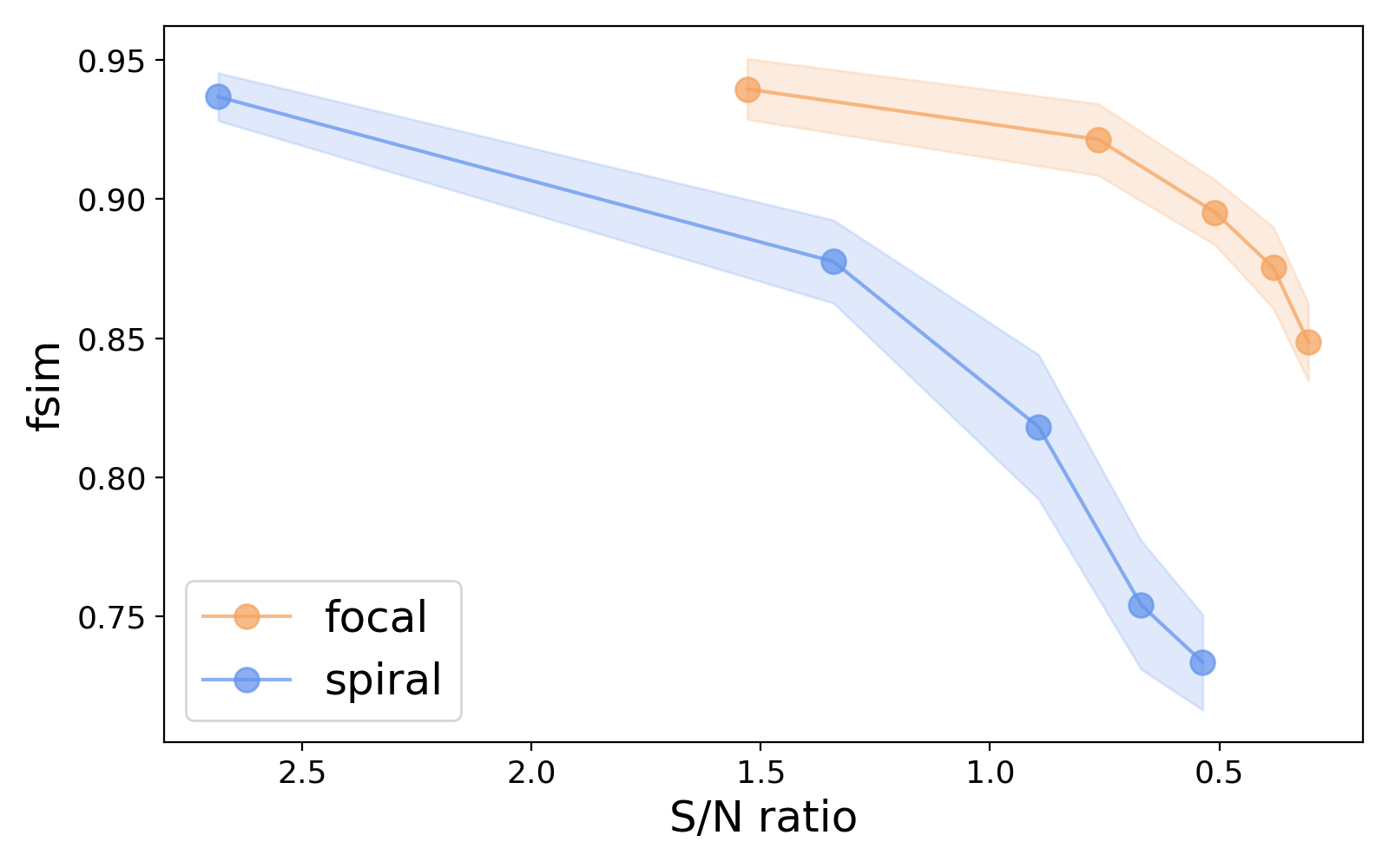}
        \caption{}
        \label{fig:noise4}
    \end{subfigure}
    \caption{Snapshots of the reconstruction of active tension waves for different levels of Gaussian noise, for 2 foci \textbf{(a)} and a rotating spiral \textbf{(b)}. Figures illustrate typical deformation components used as input data next to the estimated active tension for one time instance. Panel \textbf{(c)} shows the quality of reconstruction for increasing noise level. The same result is drawn in terms of signal-to-noise ratio in panel \textbf{(d)}. 
    \label{fig:noise} 
    }
\end{figure}

\begin{figure}[ht!] \centering
	\includegraphics[width = 0.4 \textwidth]{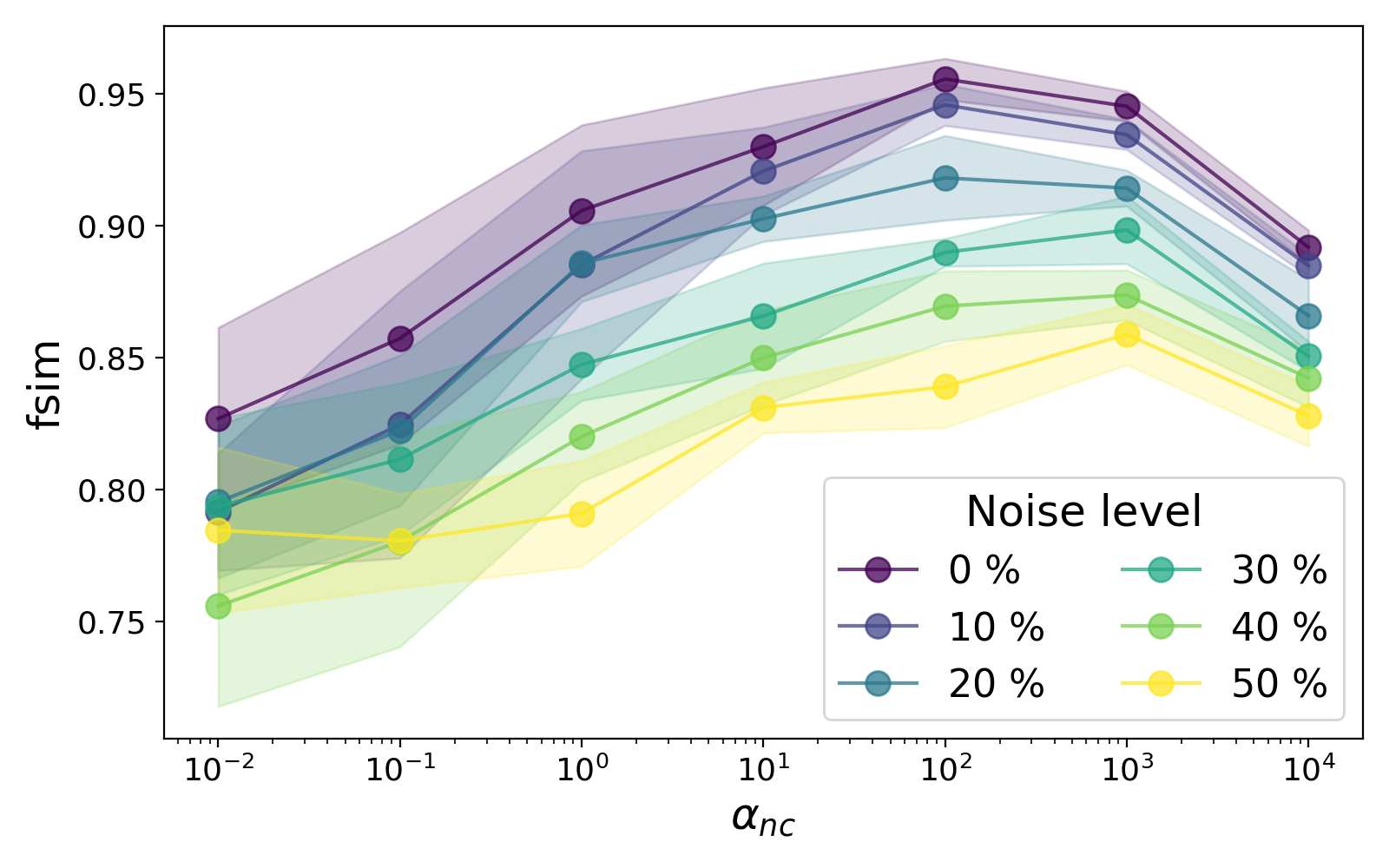}
	\caption{Fsim scores of the focal reconstruction in function of the weighting parameter $\alpha_{nc}$ which signifies the strength of the physics term, for different Gaussian noise levels. The optimal values are slightly larger for higher noise, indicating the use of a correct balance between data ($\alpha_{data}=1$) and physics.}
    \label{fig: fsim alpha nc}
\end{figure}

In this section we investigate how noise in the data set affects the reconstruction of the active tension wave. Panels (a) and (b) in Fig. \ref{fig:noise} show the noisy data in the first two columns, together with the reconstructed active tension profile, for different noise levels. Remarkably, for the focal pattern in panel (a), the overall shape can be recognized quite well even up to a noise level of $50\%$. The amount of distortion in the reconstruction stays minimal for most cases. The spiral wave depicted in panel (b) is affected more by the increasing noise levels, as its core is not well resolved anymore beyond $30\%$ noise level. To quantify the accuracy of the reconstructions, the bottom panels (c) and (d) show the time averaged fsim score in function of Gaussian noise percentage and signal-to-noise ratio. For each noise level, ten different models were constructed and optimized, differing in the randomly selected initialization of the neural network and mini-batch construction, and the Gaussian noise sampling. The graphs show, as expected, a decrease in reconstruction quality when noise becomes more present. The slope is steeper for the more complex spiral pattern in comparison to the focal data. Note that there is a non-zero standard deviation in the noiseless case, indicating the effect of randomness of the neural network initialization and optimization. In case of large noise levels, it is natural to put more weight to the physics term. Therefore, we studied the effect of noise on the optimal $\alpha_{nc}$. Fig. \ref{fig: fsim alpha nc} shows the time-averaged fsim score of the reconstruction of the focal pattern in function of $\alpha_{nc}$ for different noise levels. When the noise level is increased from 0 to 50\%, the optimal $\alpha_{nc}$ shifts from $10^2$ to $10^3$. For definiteness, we report the other results here with fixed $\alpha_{nc} = 10^2$.

\subsubsection*{Effect of spatial resolution}
\label{subsubsec: resolution}

\begin{figure}[ht!]
    \centering
    \begin{subfigure}{0.35\textwidth}
        \includegraphics[width=\linewidth]{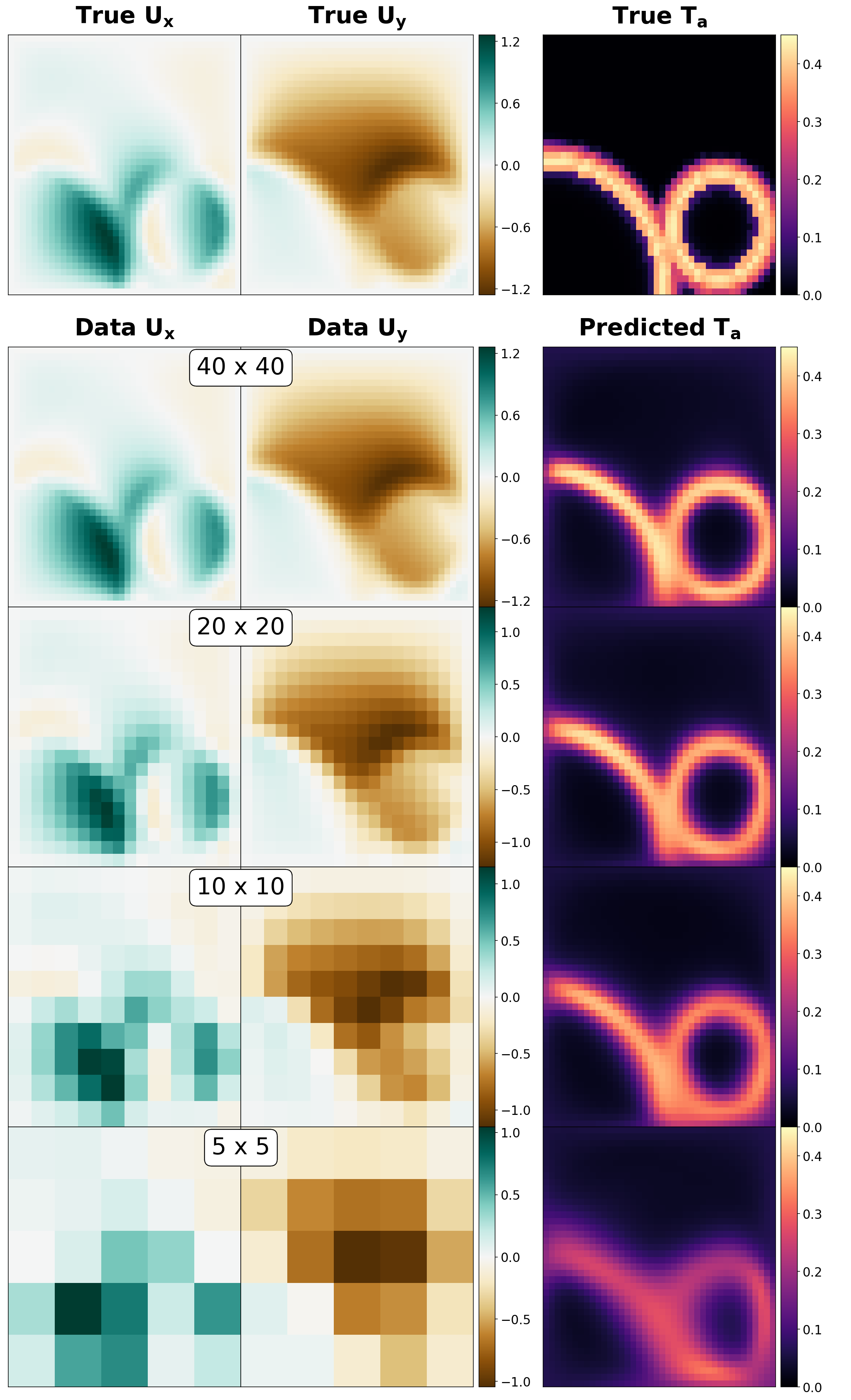}
        \caption{}
        \label{fig:resolution-a}
    \end{subfigure}
    \hspace{0.05\textwidth} 
    \begin{subfigure}{0.35\textwidth}
        \includegraphics[width=\linewidth]{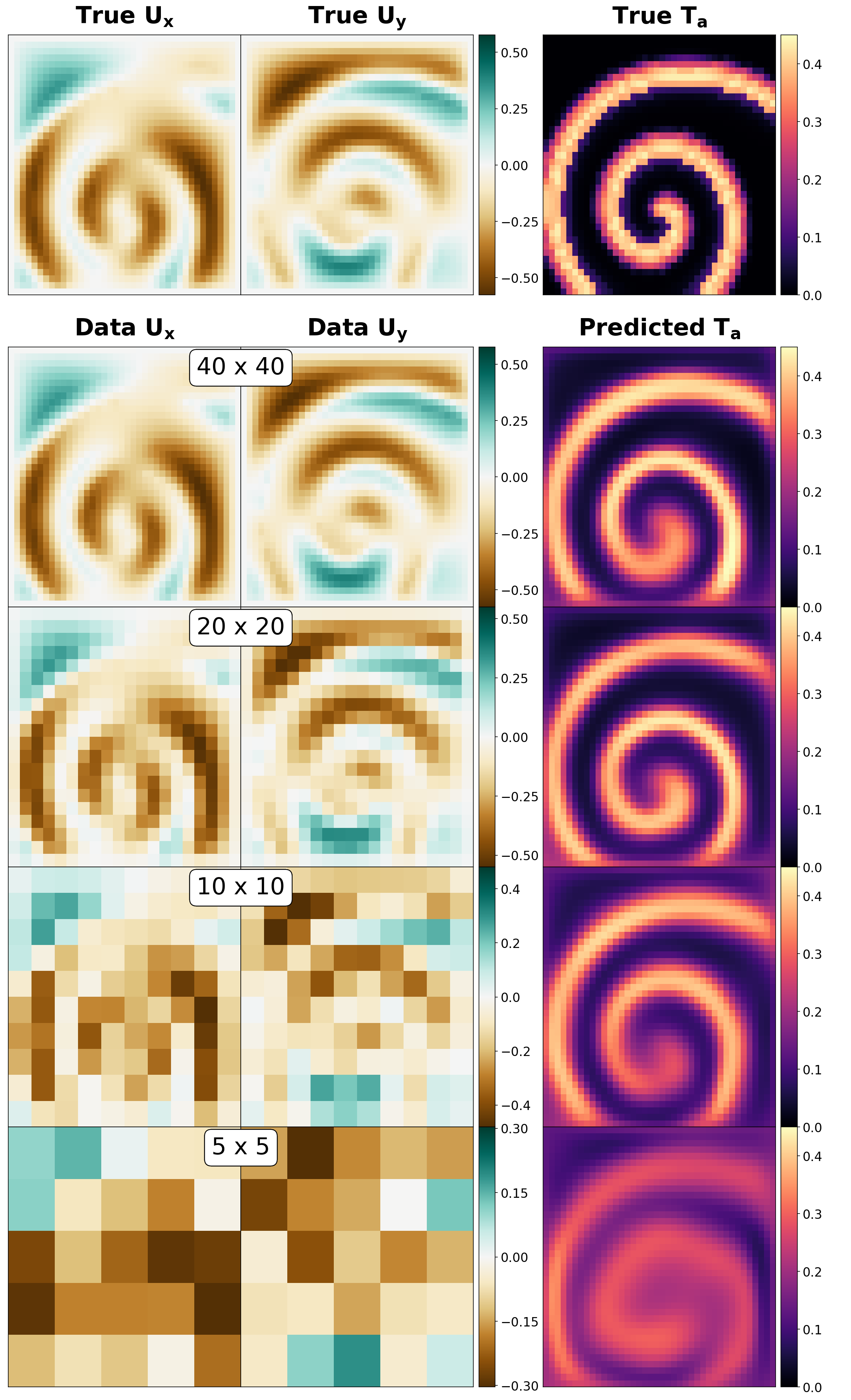}
        \caption{}
        \label{fig:resolution-b}
    \end{subfigure}
    \begin{subfigure}{0.35\textwidth}
        \includegraphics[width=\linewidth]{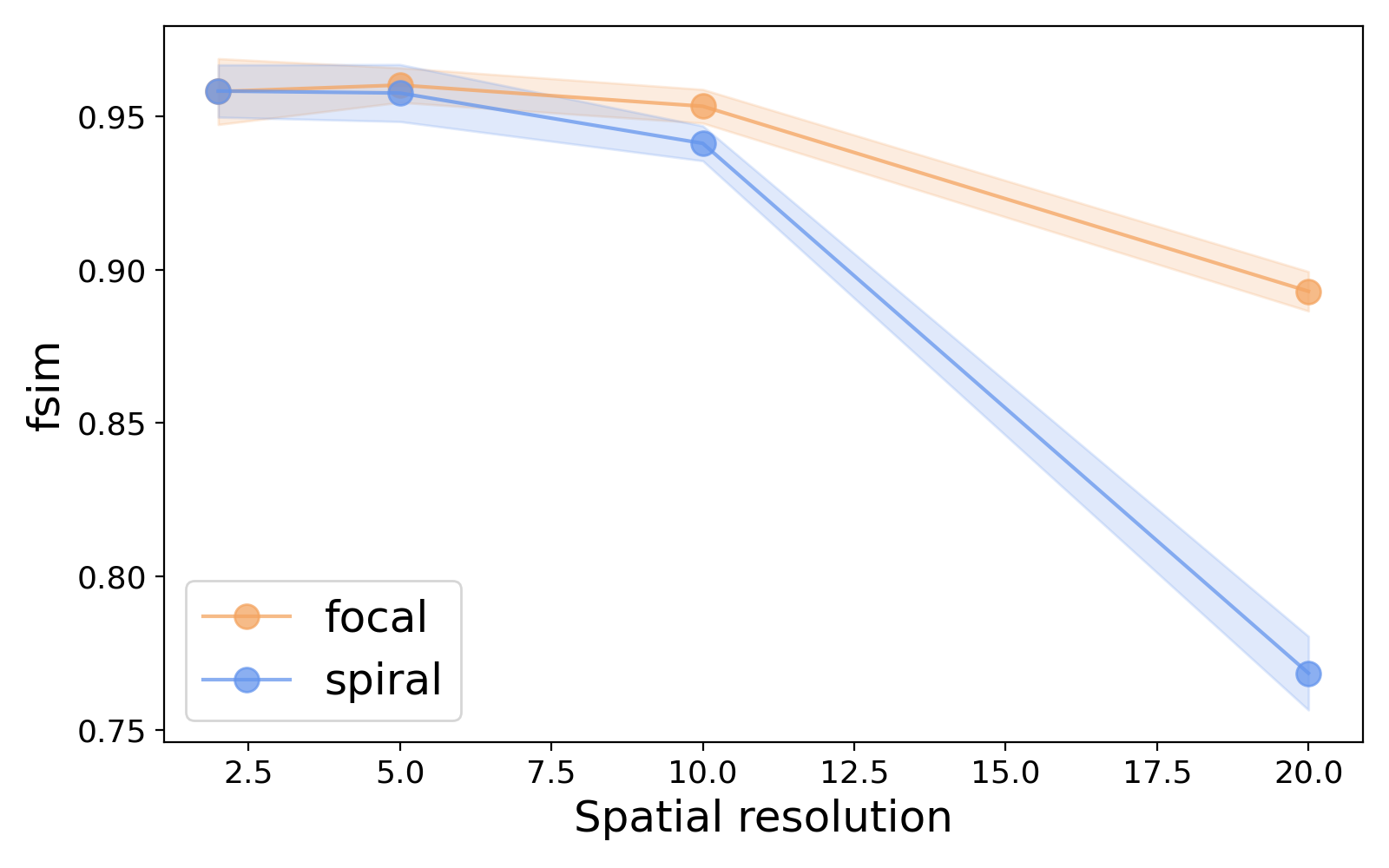}
        \caption{}
        \label{fig:resolution-c}
    \end{subfigure}
    
    \caption{Reconstructions of the active tension wave for different spatial resolutions. Panels \textbf{(a)} and \textbf{(b)} illustrate the deformation components used as data in the optimization, as well as the resulting $T_a$ prediction. The focal and spiral pattern are well predicted up to resolutions of $10$, while main features are still visible for resolutions of $20$ (5x5 data points). Panel \textbf{(c)} shows fsim scores in function of spatial resolution. Averages and standard deviations are presented for 10 randomly initialised models.}
    \label{fig:resolution}
\end{figure}

In addition to investigating noise effects, we varied the spatial resolution of the deformation data used as input for the PINN. Ten models were optimized with different random initialization as before. Panels (a) and (b) in Fig. \ref{fig:resolution} show the active tension reconstruction for one time instance, while the resolution of the deformation data is decreased. Above a spatial resolution of $10$ there is almost no difference from the fully resolved reference case, both for the focal and spiral pattern, while the shape and magnitude are accurately captured. If we downsample the resolution to $20$, the data square is 5 by 5 pixels, but the main pattern is still visible even though the sharp boundaries are not resolved anymore. This trend is also clearly shown in the feature similarity graph in panel (c), which  indicates a significant drop once the resolution gets larger than $10$. The decrease in reconstruction quality is smaller for the focal than for the spiral case.

\subsubsection*{Tri-planar projections}
\label{subsubsec: projected slices}

\begin{figure}[ht!]
    \centering
    \begin{subfigure}{0.49\textwidth}
        \includegraphics[width=\linewidth]{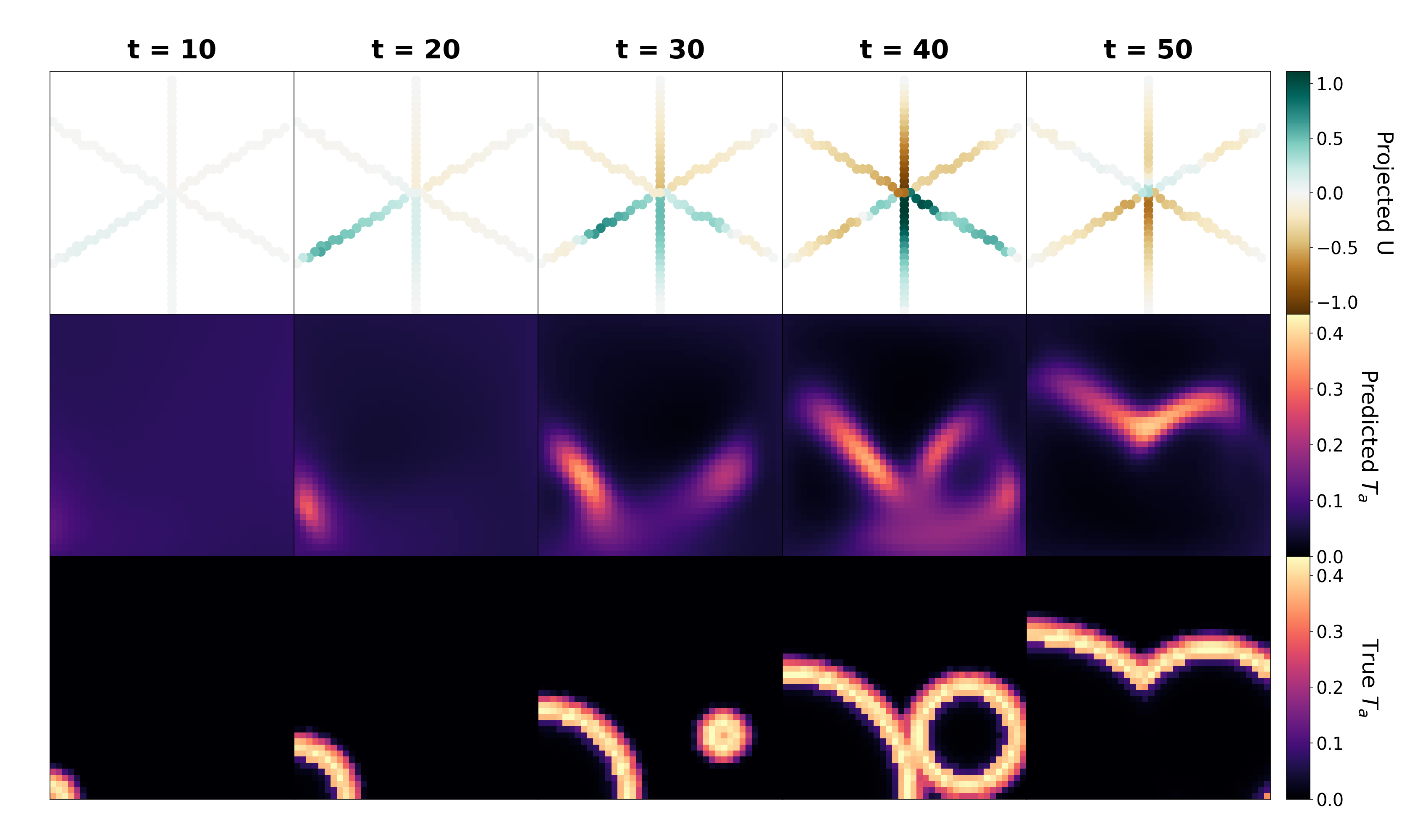}
        \caption{}
        \label{fig:slices-a}
    \end{subfigure}
    \hspace{0.01\textwidth}
    \begin{subfigure}{0.49\textwidth}
        \includegraphics[width=\linewidth]{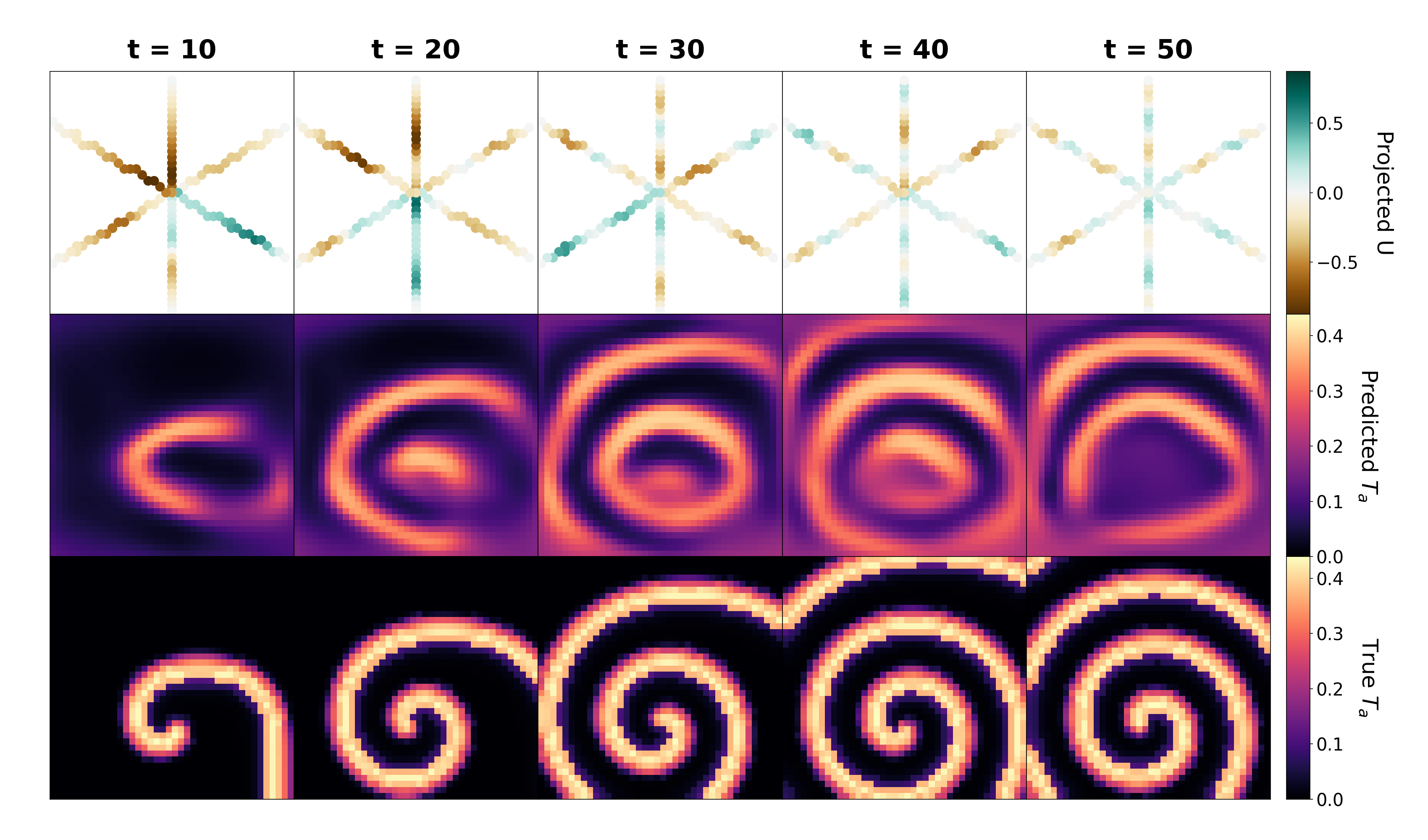}
        \caption{}
        \label{fig:slices-b}
    \end{subfigure} 
    \caption{Reconstructions of the active tension wave for projected tri-planar data using the NC-PINN (Fig. \ref{fig:standard-pinn}), for the focal pattern \textbf{(a)} and spiral pattern \textbf{(b)}. Only the deformation component along the projection line is used for the reconstruction.}
    \label{fig:slices}
\end{figure}

Lastly, we reduced the quality of the input deformation data even further by only keeping three 1D slices from the full 2D data space, see Fig. \ref{fig:slices}. This case resembles an example of a tri-planar acquisition and leads to very sparse data. Additionally, we only retain the deformation component along the line, which introduces more uncertainty. The network itself still predicts the full vector field $\vec{U}(x,y,t)$, as this is necessary for the physical equations, while the data term is now calculated by first projecting the predicted deformation vector on the line. Fig. \ref{fig:slices} shows reconstructed activation sequences after optimizing on three equally distant slices each containing 40 data points. For the focal pattern, the second row of panel (a) shows that the global dynamics can still be discerned: one focal source originating on the left bottom, followed by a second source on the right. However, sharp features are missing as well as parts of the wavefront close to the spatial borders ($t=50$) and at early times ($t<10$). As the spiral pattern in panel (b) is more spatially distributed, a better global reconstruction is achieved compared to the focal case. Due to the lack of sufficient information, however, it is difficult to judge whether the estimated source is a spiral or a focus.

\subsection*{Inversion using wave propagation}
\label{subsec: inversion using wave propagation}

\begin{figure}[ht!] \centering
	\includegraphics[width = 0.5\textwidth]{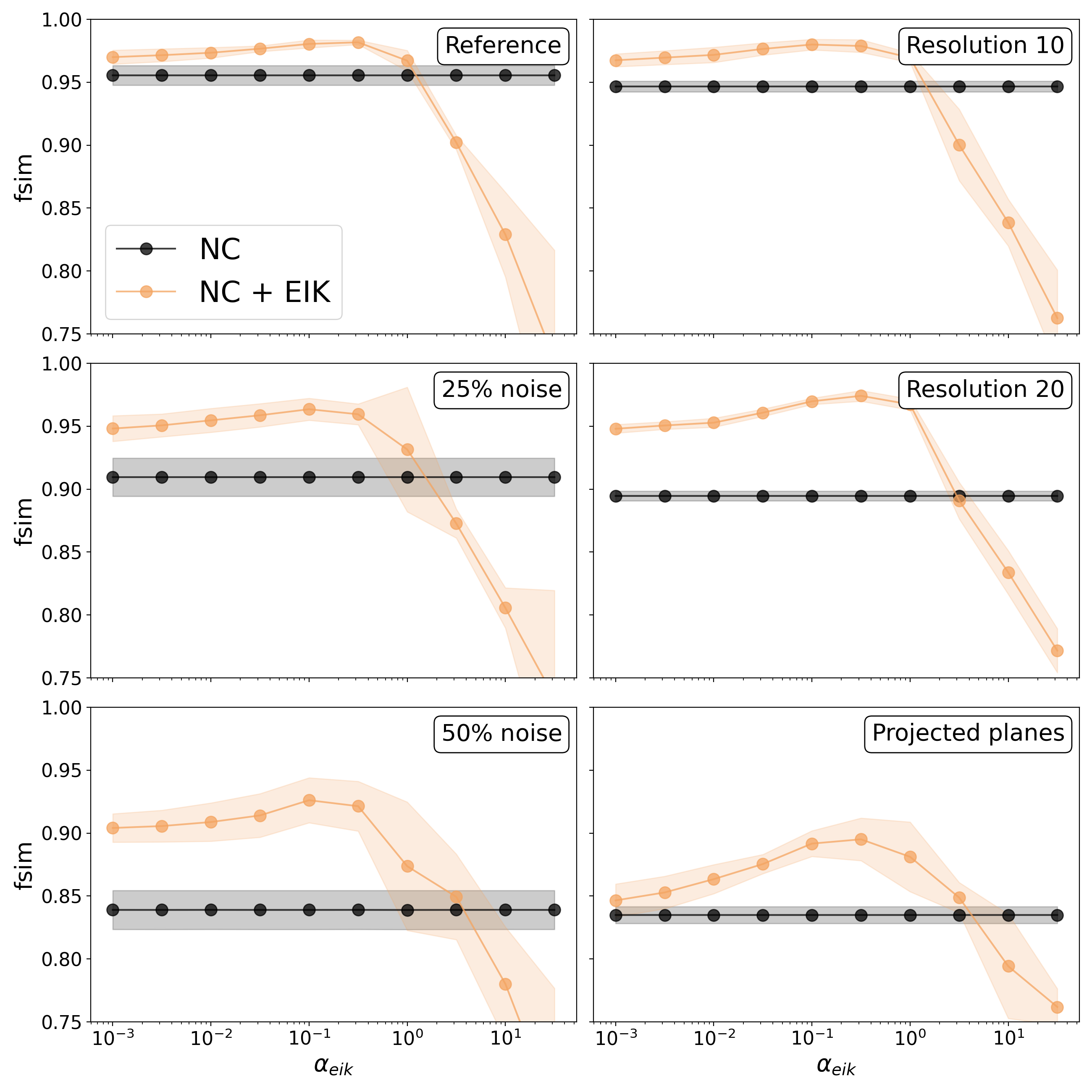}
	\caption{Fsim scores of the reconstructed focal pattern in function of the physics weighting parameter $\alpha_{eik}$. Averages and standard deviations are shown for 5 differently initiated runs by randomizing the network initialization, mini-batch construction and Gaussian noise if applicable. The different panels show the reference case (all available data), two cases of increasing Gaussian noise levels, two cases of spatially reduced data and the projected tri-planar data. All curves exhibit the same general shape: converging values towards small $\alpha_{eik}$ and declining values for large $\alpha_{eik}$, having an optimal value in between.}
    \label{fig: extended_fsims}
\end{figure}

The results in the previous section show that the PINN solely based on elastic equations may be insufficient to recover the activation pattern in very noisy, downsampled or sparse data regimes. For non-repeated activation patterns, we can improve the reconstruction with the extended NC+EIK-PINN framework, see Methods section. The second optimization step, involving the local activation time, requires to set the weighing parameter $\alpha_{eik}$ of the physical loss term, similar to $\alpha_{nc}$ for the NC-PINN. To explore its influence on the accuracy of the $T_a$-reconstruction, we kept $\alpha_{data-\tau} = 10^{-3}$ constant and varied $\alpha_{eik}$ on a logarithmic scale as only the relative weight is of importance. Each inversion was performed five times with different initialisation of the neural network, mini-batch construction and random seeds of the added Gaussian noise, if applicable. Fig. \ref{fig: extended_fsims} compares the quality of reconstruction using only the NC-PINN to the two-step approach with wave regularization. We present the reference case (no noise, full resolution), two levels of Gaussian noise, two reductions in spatial resolution and the projected tir-planar data. All cases show a similar pattern: for large $\alpha_{eik}$ fsim scores are rapidly declining and lie below the NC results since the system only takes into account the wave propagation. For small values of $\alpha_{eik}$ the regularization is absent and results approach values either above or close to the NC results. In the intermediate region, however, we find maximal fsim scores, indicating parameter values for which the combined approach reaches its full potential.\\

\begin{figure}[ht!] \centering
	\includegraphics[width = 0.6\textwidth]{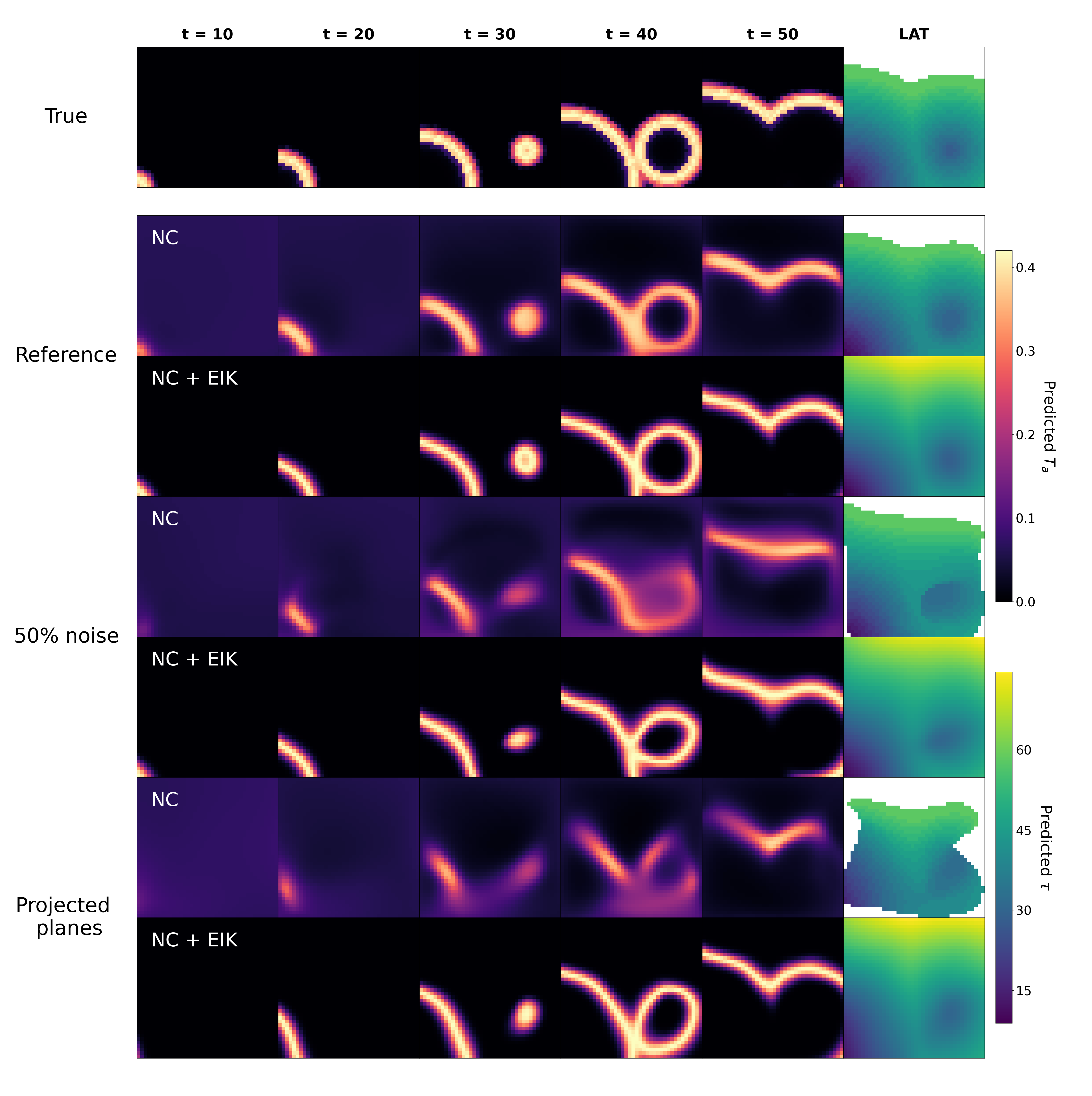}
	\caption{Comparison of the reconstructions of the active tension wave using either the NC-PINN or the combined approach (NC + EIK). The first row presents snapshots of the true $T_a$-field together with the calculated LAT map. Rows below show the results for the reference case (ideal data), high noise level of $50\%$ and the projected tri-planar data. If the quality of the inital reconstruction (NC) is lower, the second optimization (EIK) has more potential to improve the result as can be seen through the LAT extrapolation in the tri-planar case.}
    \label{fig: extended_t_series}
\end{figure}

Fig. \ref{fig: extended_t_series} compares the reconstructed active tension waves via both methods to the ground truth in the top row. For the combined approach, we used $\alpha_{eik} = 0.1$ and $\alpha_{\tau}=10^{-3}$. The number of layers and nodes per layer were chosen as $N_l'=5$ and $N_n'=10$ in order to have sufficient degrees of freedom to capture the simpler 2D solution space ($\tau$($x,y$)). Computational times for the second optimization step were measured to be $26.24\pm0.61$\,s ($10$ independent runs). In the reference case i.e. no noise and full spatial sampling, the combined method gives very similar results to the one-step method. The improvement is minimal as the original pattern is already an accurate reconstruction. For the case with extreme noise, the original map of local activation times $\hat{\tau}$ is quite complete, such that the second optimization step is mainly interpolating and overwriting incorrect $\tau$-data points. Also, due to the prescribed Gaussian profile of fixed amplitude used for the final $T_a$ reconstruction, the contrast in the combined method is higher. The third inversion problem with projected tri-planar data clearly shows the advantage of the combined method. Here, the original $\tau$-map is incomplete and more extrapolation needs to be done. This can also be seen in Fig. \ref{fig: extended_fsims} for which the noise and resolution cases, the reconstruction already benefits from the Gaussian wave pattern on top of the $\tau$-map without significant influence of the eikonal equation. The results based on the projected planes data, however, show the necessity for the additional physics loss term. In conclusion, we find an overall improvement using the combined method and this improvement increases if the input data is of lower quality.

\section*{Discussion}
\label{sec: discussion}


This work presents a potential methodology to non-invasively measure the cardiac electrical activation pattern by utilizing mechanical deformation data. To solve this complex inversion task, we proposed the use of physics-informed neural networks which combine the strength of recent breakthroughs in machine learning with the explicit physical knowledge of the system. Within this study, we used fully connected neural networks to estimate the active tension waves from noisy and sparse deformation data obtained from simulations of a 2D isotropic medium with linear elasticity. Both focal and spiral patterns could be accurately reconstructed in a short period of time i.e minutes on a general-purpose laptop even without potential GPU parallelization.\\

In contrast to supervised machine learning methodologies, PINNs do not require large data sets to train on. This can be particularly useful in complex medical problems, in which quality data to train the model on is often limited or completely absent. Training data is then acquired by running large-scale simulations with specialized software, while covering as much of the parameter space as possible. Combining this with the extensive training time of the model itself, computational costs are substantially higher compared to PINNs, at least in the first stage of constructing the model. Additionally, by enforcing the physical laws directly on one instance, we can easily control and interpret the results as well as insert patient-specific information that would be relevant e.g. impose scar positions from previous operations. If important information is missing, however, supervised methods are  still able to generate a first estimate more easily as they do not require specific models or parameter sets during evaluation. We expect that hybrid models and strategies could in the near future benefit from the advantages of both approaches.\\

In this work, a global optimization strategy via physics-informed neural networks has allowed us to reconstruct the active tension from deformation data that has significant noise or lacks good spatial resolution. Going one step further, we showed how easily the network could be adapted to only include certain deformation components. More specifically, we optimized PINNs on deformation data projected on three 1D-planes, $60\degree$ apart from each other. We found that the most important features of the patterns could still be captured, even with the extreme sparseness and reduction of total information. As such, our next effort will be to adapt the method to three-dimensional cardiac geometries, after which full 3D or tri-planar echo data\cite{Ramalli2020} may be used to recover the activation sequence.\\ 

In situations where data is either extremely noisy, coarse or sparse, wave propagation properties can be used in order to further regularize the problem. More specifically, we introduced a second PINN which was constructed in the reduced LAT space to include both information on the propagation speed (eikonal loss term) and wave shape (Gaussian shape in time). Clear improvements, as evaluated by the fsim score, were shown for all data sets, with greater upside for the worst-quality data sets. It is interesting to note that in the sparse, tri-planar case, the wave regularization was necessary to obtain improved reconstructions, as the Gaussian shape alone had little influence on the accuracy. However, to achieve these results, additional information is needed on the expected active tension wave properties i.e. the wavelength or duration of the profile and velocity of the wave. It is possible to set these values within physiological regimes or to estimate them during the PINN optimization, but this task was outside our current scope. Future work could include studying the effect of parameter misfits on the excitation waves. An important question in view of applications is how accurate the prediction should be in specific medical situations and if the wave propagation improvement is necessary or redundant. If for example, LAT maps serve as a first approximation to estimate the origin of an ectopic beat, the first optimization step could be sufficient in most cases. If, however, more precise reconstructions are necessary in either $\tau$- or $T_a$-space, additional wave information can be valuable. Note that the increase in computation time is minimal as in our experiments the second optimization part was performed $8$ times faster, due to the smaller architecture of the second PINN, the smaller amount of physical equations and the absence of higher order derivatives.\\

PINNs are optimized through the minimization of a single loss function consisting of two or more loss terms. As in any optimization task, the balance between data and regularization is crucial to obtain satisfactory results. In this study, the weighting parameters $\alpha_*$ served this role, and we have investigated their influence on the reconstruction for both the elastic and eikonal regularization. More specifically, optimal results were achieved when $\frac{\alpha_{nc}}{\alpha_{data}}$ and $\frac{\alpha_{eik}}{\alpha_{data-\tau}}$ had values of around $10^{2}$ and $10^{-3}$-$10^{-2}$, respectively. In practical applications, we cannot make this choice on the basis of an external measure of accuracy as no true solution is known and we need to resort towards prior knowledge of similar in silico or other experiments. A good starting point, however, can be a physical rescaling of the loss terms, if such information is available. This approach will put the different loss terms, at least approximately, on the same foot, while also making the $\alpha_*$ choices independent of the relevant physical scales. Table \ref{tab: scaling} shows for each $\alpha_*$ used in this work, its implicit dimension and proposed scaling factor. For the two data terms, we make use of the maximal values of either the data itself ($U_ {x,max}$, $U_ {y,max}$) or the temporal domain of the system ($\tau_{max} = T_{max}$). The factors for the physics terms are based upon characteristic scales including certain parameters. Even though their exact values are in general not known, approximate values can still serve that goal. If we apply these scaling factors on the optimal $\alpha_*$ values found earlier, we notice in Table \ref{tab: scaling} the same order of magnitude for $\alpha_{data}$ and $\alpha_{nc}$. It indicates that, after rescaling, a balanced loss function is beneficial. In the second PINN, the rescaling results in an unbalanced weighting ratio $\frac{\alpha_{eik}}{\alpha_{\tau}}$ of $10^{-2}$-$10^{-1}$. This could be explained as following: In the NC-PINN, the physics term involves, next to $\vec{U}$, the $T_a$-field which is not heavily constrained in other ways. In contrast, the eikonal term in the EIK-PINN only depends on the $\tau$-field, resulting in a competitive scenario wherein both terms directly influence the same variable ($\tau$). Additionally, in contrast to the Navier-Cauchy equation which fits the data in the full spatiotemporal domain (except the boundaries), the eikonal equation does not. Both source positions as well as regions where two wavefronts meet, have a much smaller spatial gradient than expected. The combination of these arguments could explain the larger data weighing factor in the total loss function for the optimal EIK-PINN. Another study\cite{Rohrhofer2023} came to similar conclusions, noting that in the case of physically scaled loss terms, a larger data term was beneficial towards solving partial differential equations.\\

\begin{table}[ht!]
\centering
\resizebox{0.6\columnwidth}{!}{
\begin{tabular}{|c|c|c|c|c|}
\hline
\multicolumn{1}{|c|}{\textbf{$\alpha_{*}$}} & \multicolumn{1}{c|}{\textbf{Optimal $\alpha_{*}$}} & \multicolumn{1}{c|}{\textbf{Dimensions of $\alpha_{*}$}} & \multicolumn{1}{c|}{\textbf{Scaling factor}} & \multicolumn{1}{c|}{\textbf{Rescaled optimal $\alpha_{*}$}}\\
\hline
\hline
$\alpha_{data}$ & $10^0$ & $L^{-2}$ &${U_ {x,max}}^{2}, {U_{y,max}}^{2}\approx 1$  & $1$ \\
$\alpha_{nc}$& $10^2$& $(\frac{L}{P})^2$ &$(\frac{T_{a,max}}{v\cdot\sigma})^2 \approx \frac{1}{80}$ & $1.25$ \\
\hline
$\alpha_{data-\tau}$& $10^{-3}$ & $T^{-2}$ &${\tau_ {max}}^{2}\approx 3000$ & $3$ \\
$\alpha_{eik}$ &$10^{-1}-10^{0}$& $(\frac{L}{T})^2$ &$v^{-2}\approx 0.4$ & $0.04-0.4$ \\
\hline
\end{tabular}}
\caption{\label{tab: scaling} Estimates of the required magnitude of weighing parameters by non-dimensionalising the loss terms. Dimensions refer to length (L), time (T) or pressure (P). 
}
\end{table}

This study aimed to demonstrate the feasibility of non-invasive arrhythmia imaging, by casting it as a global optimization problem and relying on physical principles. Our proof-of-concept study consisted of a simplified setting. First, the domain was a 2D regular square, enabling the use of continuous collocation points, while also reducing the computational burden. Second, the synthetic data used in this work was generated by a simple isotropic mechanical model with linear elasticity. In this way, deformations were relatively small with respect to the spatial coordinates so that we could work in the same reference frame for all time steps. Another limitation of this study is the prior knowledge of the 'correct' underlying physical model together with the model parameters. Note here that the PINN architecture does not discriminate between estimating physical parameters (e.g. conduction velocity or material constants) and neural network parameters. This seamless equality allows to optimize physical parameters alongside the prediction of the field variables. To what extent the exact model and physical parameters need to be known or can be estimated during the optimization in order to have sufficient reconstructions, will be the subject of further research. Lastly, as the underlying activation was the focus of this study, we only reconstructed the active tension. Including the electro-mechanical coupling between $T_a$ and the local transmembrane voltage would be a straightforward extension and could be useful in more complicated cases where there is heterogeneous coupling or when important electrical features (e.g. arrhythmia substrates) cannot be observed in the mechanical waves.\\ 

In conclusion, we reconstructed active tension waves from spatiotemporal recordings of deformation data by leveraging the strength of physics-informed neural networks. 2D synthetic data consisted of both focal and spiral activation waves, causing mechanical elastic response via the isotropic Navier-Cauchy equations. By means of the feature similarity index, we showed that accurate reconstructions are possible even with large Gaussian noise levels, reduction of spatial resolution and tri-planar, projected data. Additionally, by incorporating information about the wave propagation, improvements in the reconstruction could be obtained. This was especially relevant when the data quality was heavily reduced such as the sparse tri-planar case. Further research will focus on more complex non-linear coupled problems in cardiac 3D geometries. We here provide evidence that PINNs have the potential to thrive in future clinical settings, as they encompass and interesting mix of recent advances in machine learning, optimization and explicit physical knowledge. The PINN formalism may open up pathways to recover spatiotemporal activation sequences from non-invasive ultrasound data obtained in patients.

\bibliography{main}

\section*{Acknowledgements}

The authors are grateful for helpful discussions to the following colleagues: Jan D'hooge, Joris Ector, Desmond Kabus and Konstantina Papangeloupoulou. 

\section*{Funding}
This research was funded by KU Leuven grant C24E/21/031.

\section*{Author contributions statement}

H.D. conceived the concept of the study. N.D. implemented the code, ran the numerical experiments and interpreted the data. H.D. and N.D. together designed the algorithms, interpreted the results and wrote the paper. All authors reviewed the manuscript. 

\section*{Additional information}

The authors declare no competing interests. 

\end{document}